\documentclass{aa}  

\usepackage{graphicx}
\usepackage{txfonts}
\usepackage[colorlinks=true,
     linkcolor=blue,
     filecolor=blue,
     citecolor = blue,      
     urlcolor=blue,]{hyperref}
\usepackage{graphicx}	
\usepackage{amsmath}	
\usepackage{amssymb}	
\usepackage{siunitx}
\usepackage[dvipsnames]{xcolor}
\usepackage{float}
\usepackage[normalem]{ulem}

\addto\extrasenglish{}
\addto\extrasenglish{}
\addto\extrasenglish{}
\addto\extrasenglish{}
\addto\extrasenglish{}
\addto\extrasenglish{}
\addto\extrasenglish{}
\addto\extrasenglish{}
\newcommand{\aref}[1]{\hyperref[#1]{Appendix~\ref{#1}}}

\definecolor{darkgreen}{rgb}{0.13, 0.55, 0.13}

\makeatletter
\renewcommand*\aa@pageof{, page \thepage{} of \pageref*{LastPage}}
\makeatother

\newcommand{\machc}{\mathcal{M}_{\rm c}}
\newcommand{\kms}{\rm km \, s^{-1}}
\newcommand{\vc}{v_{\rm c}}
\newcommand{\cs}{c_{\rm s}}
\newcommand{\mach}{\mathcal{M}}
\newcommand{\trot}{t_{\rm rot}}
\newcommand{\rc}{R_{\rm c}}
\newcommand{\rd}{R_{\rm d}}

\begin{document}

   \title{Formation of filaments and feathers in disc galaxies: Is self-gravity enough?}


   \author{Raghav Arora
          \inst{1},
          Christoph Federrath \inst{2,3}, 
          Mark Krumholz \inst{2} , 
          \and 
          Robi Banerjee \inst{1}
          }

   \institute{Hamburger Sternwarte, Universit\"at Hamburg, Gojenbergsweg 112, 21029 Hamburg, Germany\\
              \email{raghav.arora@uni-hamburg.de}
         \and
             Research School of Astronomy and Astrophysics, The Australian National University, Canberra, ACT~2611, Australia
         \and
             Australian Research Council Centre of Excellence in All Sky Astrophysics (ASTRO3D), Canberra, ACT 2611, Australia
         \\
             }

   \date{Received xxxx; accepted xxxx}
   
\titlerunning{Filaments and feathers in disc galaxies}
\authorrunning{Arora et al.}

 
  \abstract
   {Dense filaments, also known as feathers, are kiloparsec-scale dusty features present in nearby main sequence galaxies. Distinct from the spiral arms, filaments constitute a major portion of dense gas concentration. They are expected to play an important role in star formation and are known to harbour embedded star-forming regions and H$\,$\textsc{II} regions.}
   {We explore the origin of filaments and feathers in disc galaxies via global gravitational instability.}
   {We conduct a parameter study using three-dimensional hydrodynamical simulations of isolated disc galaxies that are isothermal, self-gravitating and are initialised in equilibrium. Our galaxies are uniquely characterised by two dimensionless parameters, the Toomre $Q$ and the rotational Mach number, $\machc=\vc/\cs$ (ratio of circular velocity to sound speed). We carry out simulations covering a wide range in both parameters.}
   {We find that galaxies with $Q=1$ form filaments within a single rotation, while galaxies with $Q \geq 2$ do not, even within a couple of rotations. These filaments are kiloparsec long and are semi-regularly spaced along the azimuth of the galaxy. Their morphology, density contrast and formation timescale vary with $\machc$, with filament spacing and instability onset time both inversely proportional to $\machc$ and the density contrast increasing with $\machc$. However, filament growth rates in all $Q=1$ galaxies are $\sim 0.5 \, \Omega$, where $\Omega$ is the angular frequency. We compare the filament spacing in our simulations with the ones in JWST/Mid-Infrared Instrument and HST observations of nearby galaxies and find them to be in agreement.}
   {Our study suggests that self-gravity and rotation alone are sufficient to form filaments and feathers, even in the absence of spiral arms or magnetic fields.  The morphologies of the resulting filaments are determined primarily by $\machc$, which parametrises the importance of thermal versus rotational support.}


   \maketitle
%
\section{Introduction}
There has been mounting observational evidence in the past few years that star-forming disc galaxies harbour a complex, filamentary network of dense gas and dust \cite[][and references therein]{thilker_filamentary_NGC628_JWST_2023}. These are elongated, kiloparsec-scale, dense features known to host star-forming regions, H$\,$\textsc{II} regions and young star clusters \citep{schinnerer_pdbi_2017, williams_phangs-jwst_2022}. Not only are they found in external galaxies, but there has also been evidence of such large-scale filamentary structures in the Milky Way \citep{zucker_bones_of_MW_2015, pantaleoni_2021, kuhn_2021, veena_2021}. External galaxies, however, give us the advantage of the `birds eye view', which allows us to study filament structure and morphology. Because of their richness in dust, they are visible as attenuation features in Hubble Space Telescope (HST) images \citep{elmegreen_1980, vigne_2006}, or as emission features in the near/mid-infrared, for example in the $8 \, \mu \rm m$ infrared band of Spitzer \citep{elmegreen_2006, elmegreen_2014,elmegreen_2018, elmegreen_2019} and in recent JWST/Mid-Infrared Instrument (MIRI) observations \citep{williams_phangs-jwst_2022, thilker_filamentary_NGC628_JWST_2023, meidt_rosolowsky_23}, which have reached unprecedented resolutions of $\sim10\, \si{pc}$. This enables us for the first time to compare the detailed morphology of dense gas in simulations with observations of nearby galaxies and to test the theories for the origins of filamentary structures.

The question of their origin is far from settled, and a number of different mechanisms have been proposed. Filaments have been most commonly found in grand-design spiral galaxies, and within spirals most frequently in Sb-Sc galaxies \citep{vigne_2006}. This increased detection frequency in grand-design spirals has led some authors to posit that filament formation is driven by spiral arms, and to carry out linear stability analyses and two-dimensional (2D) simulations showing that gas traversing arms created by an external spiral potential is prone to a hydrodynamical instability that causes them to develop over-densities, which are then sheared into filamentary structures in the inter-arm regions \citep{wada_koda_2004,dobbs_bonnel_2006,lee_feathering_2012, kim_wong_kim_2014, lee_feathering_2014, kim_wong_kim_2015, sormani_sobacchi_2017, mandowara_physical_2022}. Other authors have attributed filaments to Parker instability driven by large-scale magnetic fields \citep{bastian_2019} or supernova feedback from star-forming regions in the arms \citep{kim_wong_tigress_2021}. Yet other authors suggest that gas self-gravity alone could potentially create filamentary structures without the need to invoke external potentials, supernova feedback or magnetic fields \citep{griv_stability_2012, meidt_molecular_2022}.

At present, however, we lack a systematic, 3D, global, numerical study that aims to disentangle the contribution of different physical processes to dense structure formation in galaxies. Prior works that explicitly focus on gravitational instability in a galactic disc are either 2D \citep{shetty_ostriker_2006,griv_wang_2014_hydro_sims_twoD} or are low-resolution models with limited parameter space that were unable to resolve the spatial scales necessary for filament formation \citep{kim_formation_2006,wang_equilibrium_2010}. More recent and higher resolution simulations have different aims. Some are focused on studying Milky-Way analogues that have similar mid-plane pressures, distribution of the total mass into the halo, stellar bulge and the gas components as our own Galaxy \citep{agora_project_kim_2016, jeffreson_2020}. Others include complex physical processes such as an external spiral potential and subgrid models for star formation and feedback that make it hard to separate effects of different physical processes \citep[for example;][]{tress_simulations_2020, robinson_wadsley_2024, bo_pudritz_2024, khullar_2024}.

The aim for this study is to use numerical simulations to determine the conditions of filament formation in an idealised setting where we have control over the different potential drivers of filament formation and thereby can carry out clean numerical experiments to gain insight into the underlying physics. To this end, we carry out high-resolution 3D simulations of isothermal, self-gravitating isolated disc galaxies that are initialised in equilibrium. The main question we address is whether self-gravity by itself, in the absence of the other mechanisms described above (such as magnetic fields, stellar spiral potentials, supernovae), is sufficient to create the dense filamentary network observed in galaxies. We demonstrate that it is, and conduct a parameter study to quantify the morphological features of the filamentary structures that form in our simulations. Finally, we compare the physical spacing of the simulated filaments with recent observations.

The paper is organized as follows: \autoref{sec:method} describes our simulation setup, initial conditions and the library of simulations used in this work. In \autoref{sec:results}, we discuss the basic morphology of our galaxies and the filaments that form in them. We further quantify the dependence of the morphology and filament formation timescales on the initial galaxy-wide parameters. In \autoref{sec:Discussion}, we compare our results with observations and existing theoretical and numerical work. Finally, we summarise our conclusions in \autoref{sec:conclusions}.

\section{Methods} \label{sec:method}
\subsection{Physical setup} \label{sec:simulationSetup} 
Our 3D disc galaxy simulations consist of self-gravitating isothermal gas in an analytical axisymmetric, time-steady dark matter plus stellar potential with a flat rotation curve. We initialise our disc galaxies in equilibrium using the ``potential method'' described by \citet{wang_equilibrium_2010}. Here, we briefly describe the setup. 

\subsubsection{Basic setup}

We seek to initialise our simulated galaxies in equilibrium, such that the forces due to thermal pressure, self-gravity, dark matter and rotation balance in both the radial and vertical directions. The surface density of the galaxies follows a modified exponential profile given by 
\begin{equation} \label{eqn:analytical_surfaceDensity}
    \Sigma (R) = \Sigma_{\circ} \exp{\left [ -\frac{R}{R_{\rm d}} -\beta \exp \left (-\alpha \frac{R}{R_{\rm d}}\right ) \right]},
\end{equation}
where $R$ is the galactic (cylindrical) radius, $\Sigma_\circ$ is a scaling constant that will vary from run to run, $R_{\rm d} = 3\, \si{kpc}$ is the disc scale length, and $\alpha = 2$ and $\beta = 1/2$ are constants that determine the shape of the the surface density profile near the centre of the galaxy. We use this function instead of the pure exponential in order to flatten the surface density profile close to the centre and guarantee that the scale height there is resolved well enough ($\geq4$ cells) to avoid artificial numerical oscillations due to inadequate resolution of the vertical pressure gradient that balances self-gravity. Since we are not interested in the central part of the galaxy, this does not affect our results.

The rotation profile of the disc is given by
\begin{equation}\label{equation:rotationCurve}
     v_{\rm rot} = \vc\frac{R}{\sqrt{R^{2} + R_{\rm c}^{2}}}, 
\end{equation}
where $\vc$ is the saturated circular velocity of the gas and $R_{c\rm } = 2\, \si{kpc}$. The gas follows an isothermal equation of state with 
\begin{equation} \label{eqn:isothermal_eos}
    P = \rho \cs ^{2}, 
\end{equation}
where $P$ is the gas pressure, $\rho$ is the density, and $\cs$ is the sound speed. 

We use the constraint of radial equilibrium to determine the stellar plus dark matter potential $\phi_{\rm dm}$. For the exact expression and details regarding the procedure, the reader is referred to \autoref{appendix:dark_matter_potential}. We then use this to place the disc in vertical equilibrium, where we balance the force due to the gas self-gravity and the gravity due to the stellar and dark matter components with the vertical pressure gradient of the gaseous disc at each galactocentric radius,


\begin{equation}
    \label{eq:vertical_balance}
    \rho \frac{\partial}{\partial z}\left(\phi_\mathrm{g} + \phi_\mathrm{dm}\right) + \cs^2 \frac{\partial \rho}{\partial z} = 0,
\end{equation}
where $\phi_{\mathrm{g}}$ is the gravitational potential of the self-gravitating gas (see  \autoref{appendix:dark_matter_potential}). This is subject to the constraint that 
\begin{equation}
    \label{eq:profile_constraint}
    2 \int_0^\infty \rho(R,z) \, dz = \Sigma(R),
\end{equation}
where $\Sigma(R)$ is the surface density profile (\autoref{eqn:analytical_surfaceDensity}). Since $\phi_\mathrm{g}$ and $\phi_\mathrm{dm}$ do not depend on $\rho$ in our simple approximation, we can find a vertical profile $\rho(R,z)$ that satisfies these two constraints numerically. We do this by starting with an initial guess of the midplane density $\rho(R,0)$ and integrating \autoref{eq:vertical_balance} to obtain the profile $\rho(R,z)$, which is then rescaled by a constant factor in order to satisfy \autoref{eq:profile_constraint}. We point out that this is an approximate procedure rather than exact, since the gas gravitational potential $\phi_\mathrm{g}$ should depend on $\rho$. However, we have found that the approximation is sufficiently accurate that our initial discs quickly relax to an exact equilibrium.

Finally, we immerse our galactic discs in a uniform hot circum-galatic medium (CGM) with a constant $T_{\rm CGM} = 10^7 \, \si{K}$ and density. We transition from the cold galactic disc to the hot CGM when our discs reach a fixed transition density of $\rho = 10^{-28} \,\si{cm^{-3}}$, which is $\sim 5$ orders of magnitude less than the central density of our models. The density of the CGM is such that there is no pressure gradient between the two media.

\subsubsection{Turbulent velocity field}
The rotation curve (\autoref{equation:rotationCurve}) gives the mean initial velocity of the gas, but we also impose a perturbation on top of this to seed the instabilities that we strive to study. Our procedure for doing so mirrors that described  in \cite{arora_2023}: we impose a turbulent initial velocity field with a sonic Mach number $\mathcal{M} = 0.5$ and a Kolmogorov scaling of $k^{-5/3}$ on scales $[50,\,200]\,\si{pc} $. These are close to the driving scales of turbulence of $\sim140 \pm 80\, \si{pc}$ estimated in the Milky Way \citep{chepurnov_2010}. The turbulent velocity field contains a natural mixture of solenoidal and compressible modes, and we generate it with the methods described in \citet{FederrathDuvalKlessenSchmidtMacLow2010}, using the publicly available TurbGen code \citep{FederrathEtAl2022ascl}.

\subsection{Parameter study}\label{subsect: parameter_space}
We characterise our disc galaxies with two dimensionless parameters,
\begin{align}
    \langle Q\rangle & = \frac{1}{R_\mathrm{max}}\int_0^{R_\mathrm{max}} Q(R) \, dR = \frac{1}{R_\mathrm{max}}\int_0^{R_\mathrm{max}}\frac{\kappa \cs}{\pi G \Sigma}\, dR, \label{eq:Q} \\
    \machc & = \frac{v_c}{\cs}, \label{eq:machc}
\end{align}
where $\kappa$ is the epicyclic frequency given by $\kappa^2= 4 \Omega^2 + 2R\Omega (d\Omega/dR)$, $\Omega = v_\mathrm{rot}/r$ is the angular frequency, and we take $R_\mathrm{max} = 8$ kpc. The first parameter above is the mean value of the Toomre-$Q$ parameter $Q(R)$ \citep{toomre_1964} over the inner 8 kpc of the galaxy; it quantifies the gravitational stability of our galaxies. The second parameter describes the ratio of rotation speed to sound speed in the region $R\gg R_\mathrm{c}$ where the rotation curve reaches its full speed; it relates to the rotational and thermodynamic properties of the galaxy and controls its thickness\footnote{Our $\machc$  parameter is similar to the quantity $\vc/\sigma$ used in studies of galaxy kinematics to distinguish rotation-dominated from dispersion-dominated galaxies. Note that these are not fully identical since in observations $\sigma$ contains both thermal and non-thermal contributions to the velocity dispersion. By contrast, for $\machc$ our dispersion is purely thermal.}. We show in \autoref{appendix:two_dimensionless_parameters} that these two dimensionless parameters, along with the auxiliary dimensionless ratios $R_\mathrm{max}/R_\mathrm{d}$, $R_\mathrm{c}/R_\mathrm{d}$, $\alpha$, $\beta$, and $q$ that we do not vary, fully determine both the initial state and the subsequent evolution of the system. In particular, we note that disc thickness is not an independent parameter, but is instead fully determined by $\langle Q\rangle$ and $\machc$ -- primarily the latter.

\begin{figure}
    \includegraphics[width=\linewidth]{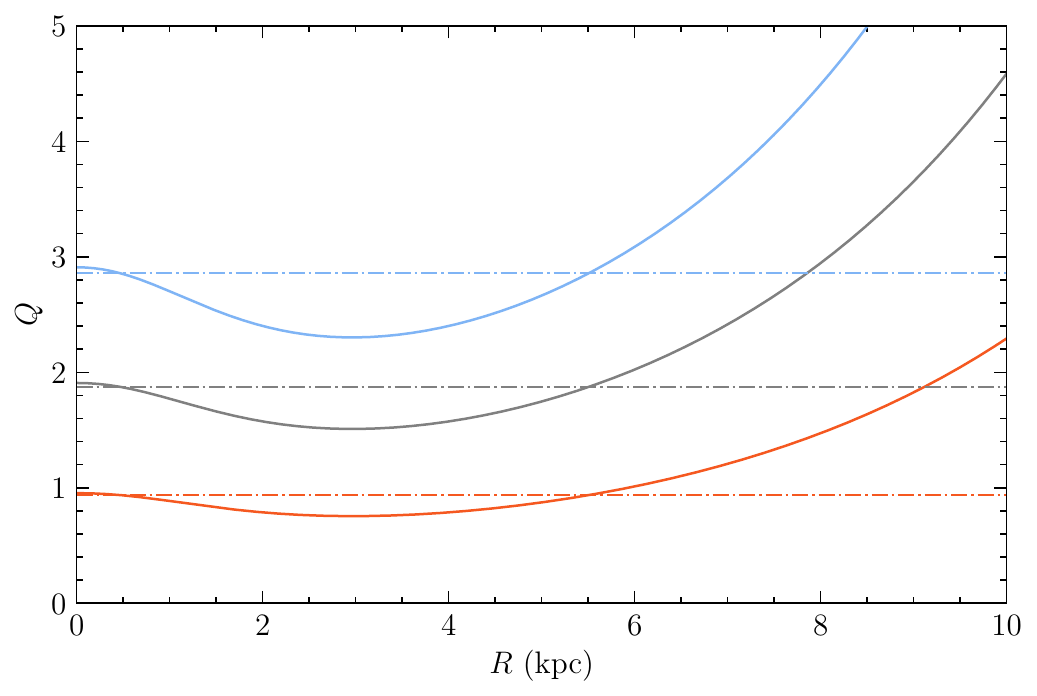}
    \caption{Initial radial Toomre-$Q$ (see \autoref{eq:Q}) profiles of our galaxies. The solid lines show $Q$ as a function of galactocentric radius and the dot-dashed lines show $\langle Q\rangle$, the average of the solid curves over the disc region $0\leq R/{\rm kpc} \leq 8$.}
    \label{fig:toomre_classical_radialProfile}
\end{figure}

The parameter space of our simulations can be split into two main groups. The first group varies Toomre-$Q$ with $\langle Q\rangle \simeq \{1,2,3\}$. (The precise values are not exactly 1, 2, and 3, and are given in \autoref{tab:initialConditions}.)
We show the radial profiles of $Q(R)$ for these three cases in \autoref{fig:toomre_classical_radialProfile}. The solid lines show $Q$ at the specified galactocentric radius and the dot-dashed lines indicate the average $\langle Q\rangle$ in the region $0 \leq R/\textrm{kpc}\leq 8$. The second group has constant $\langle Q\rangle = 1$, but varies $\machc$ with $\machc \simeq \{14, 21, 29, 36\}$; in terms of dimensional parameters, we do this by varying $\vc$ while keeping $\cs = 7$ km s$^{-1}$ fixed. The range of $\vc$ is chosen to reflect the range of rotation speeds observed in galaxies, with $\vc \in \{100, 150, 200, 250\}\,\kms$ \citep{levy_edge-califa_2018, lang_phangs_rotation_2020}. Our full parameter space is summarised in \autoref{tab:initialConditions}. For the sake of brevity, we refer to the different runs using the rounded-off values of the dimensionless parameters in the rest of the work.

\subsection{Numerical methods} \label{subsect:numerical_methods}

We use the \textsc{flash} hydrodynamical code for performing the simulations \citep{FryxellEtAl2000,dubey_fisher_2008}. The disc is initialised at the centre of a cuboidal box with side length $L_{\rm x} = L_{\rm y} = 20$~kpc in the plane of the disc and $L_{\rm z} = 2.5$~kpc in the direction perpendicular to it.  We use outflow boundary conditions as in earlier works \citep{bastian_2019, arora_2023}, but the region we use for our analysis is limited to $2 \leq R\, (\si{kpc})\leq 4$ and $H\leq 250 \, \si{pc}$ for our thickest galaxy, far enough from the edges of our box that the boundaries have negligible effects. 

We use adaptive mesh refinement (AMR) for our simulations.  Our base grid is $256\times256\times32$~cells, giving a cell size $\approx 80$~pc. However, we achieve a maximum effective resolution of $4096\times4096\times512, $ cells corresponding to a minimum cell size of $\approx 4.8\,$pc, by adding 4 additional levels of refinement. We apply this extra refinement to the same geometric region in all the simulations: a cylinder whose centre coincides with the centre of the galaxy and with radius $R_{\rm cyl} = 6 \, \si{kpc}$ and half-height $|H_{\rm cyl}| =100 \, \si{pc}$ centred on the galactic plane. This ensures that all the simulations have identical spatial resolutions. For our thinnest disc, at this resolution we have $\geq 4 $ cells per galactic scale height at galactic centre, and $\geq 8$~cells per scale height for our region of interest. Outside the cylindrical region where we enforce the highest refinement level, we refine and coarsen based on the condition that the Jeans length be resolved by at least $32$ and not more than $64$~grid cells \citep{FederrathSurSchleicherBanerjeeKlessen2011}.
 
\begin{table}
\setlength{\tabcolsep}{3.5pt} 
\centering
\caption{Parameter space of our simulations.}
\begin{tabular}{cccccccc} 
\hline 
\hline
 & Model Name             & $\langle Q \rangle $ & $\machc$ & $\vc$ & $\cs $ & $\Sigma_{\circ}$  \\
 & & & & ($\kms$) & ($\kms$)& $(\rm M_{\odot} \, pc^{-2})$ \\
\hline
1     & $Q1\_\mach36$       & 0.94   & 35.7         & 250     & 7 &  224 \\
2     & $Q1\_\mach29$        & 0.94   & 28.6         & 200    & 7 &  179  \\
3     & $Q1\_\mach21$        & 0.94   & 21.4         & 150    & 7 &  134   \\
4     & $Q1\_\mach14$        & 0.94   & 14.3         & 100    & 7 &  89.5 \\
5     & $Q2\_\mach29$        & 1.87   & 28.6         & 200    & 7 &  89.5      \\
6     & $Q3\_\mach29$        & 2.86   & 28.6         & 200    & 7 &  58.4  \\
\hline
\hline
\label{tab:initialConditions}
\end{tabular}
\tablefoot{The value of $Q$ used to parameterise the simulations is the radial average of the initial analytical profile from $R = 0\, \si{kpc}$ to $R = 8 \, \si{kpc}$.}
\end{table}

\section{Results} \label{sec:results}

\subsection{Disc morphology} \label{subsect:discMorphology}

\begin{figure*}
    \centering
     \includegraphics[width=0.99\linewidth]{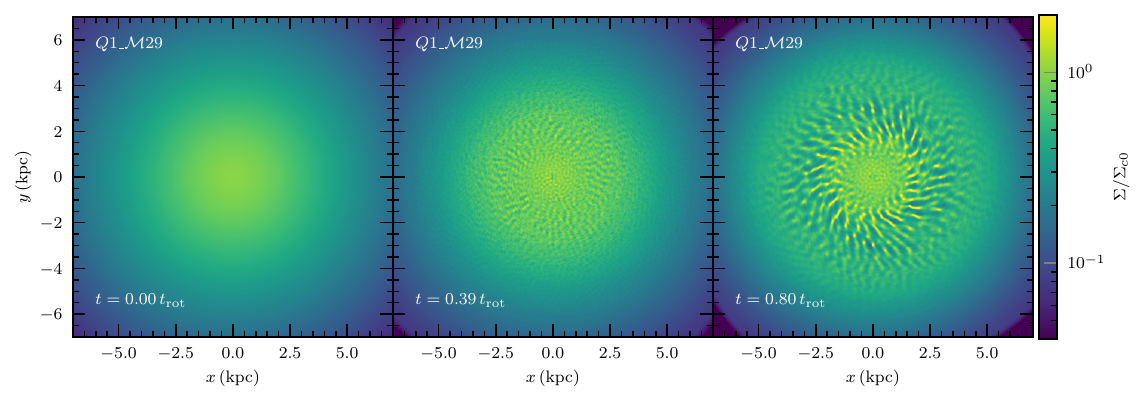}
    \caption{Projected density of the $Q1\_\mach29$ model in the z-plane. The projected density is normalised by the initial central projected density of the galaxy, shown at 0, 0.4, and 0.8~rotational times ($\trot$), from left to right. We see the gradual emergence of filamentary structures around $R=2$-$4\,\si{kpc}$.}
    \label{fig:projection_Q1_mach29_evolution}
\end{figure*}

We begin our analysis by showing the time evolution of one example run, $Q1\_\mach29$ (that is, $\langle Q\rangle \simeq 1$ and $\machc=29$), in \autoref{fig:projection_Q1_mach29_evolution}. Each panel depicts a time snapshot of the projected density of the disc in the $z$-plane. Time is expressed in units of $\trot = 2\pi R/v_\mathrm{rot}$, calculated at $R=3\,\si{kpc}$, increasing from left to right. The colourbar is normalised to the initial central surface density of the disc, that is, $\Sigma_{c0}=\Sigma(R=0, t=0)$. Following the evolution of the galaxy, we can see that it forms dense structures in less than one rotation period. These structures are $\sim\mbox{kiloparsec}$ long, uniformly spaced in azimuth, and pitched back as a result of differential rotation. For reference, we will refer to these structures interchangeably as feathers or filaments for the rest of the work. They are most prominent at radii $R\sim 2$-$4\,\si{kpc}$, which  \autoref{fig:toomre_classical_radialProfile} shows is the region of maximum Toomre instability, where $Q$ dips well below unity. These features develop even in the absence of magnetic fields, spiral arms or feedback from stars, establishing that self-gravity alone is sufficient to produce them.

Interestingly, we do not see the development of feathers in the region $R\leq 2$~kpc (at least up to the time we have run the simulations) even though it also has values of $Q\leq 1$. One possible explanation for why not is the difference in the amount of shear in the rotation curve at these smaller radii. We can quantify this through the shear parameter $q = -(d\ln \Omega/d\ln R)$, which is equal to zero for solid body rotation and one for a flat rotation curve. The value of $q$ only rises to a value of $0.5$ at $R = R_{\rm c} = 2$~kpc, and thus the lack of feathers at smaller radii is potentially a result of the low shear in this region.


\subsubsection{Dependence on $\langle Q\rangle$} \label{subsect: morphology_dependence_on_Q}

\begin{figure*}
    \centering
     \includegraphics[width=0.99\linewidth]{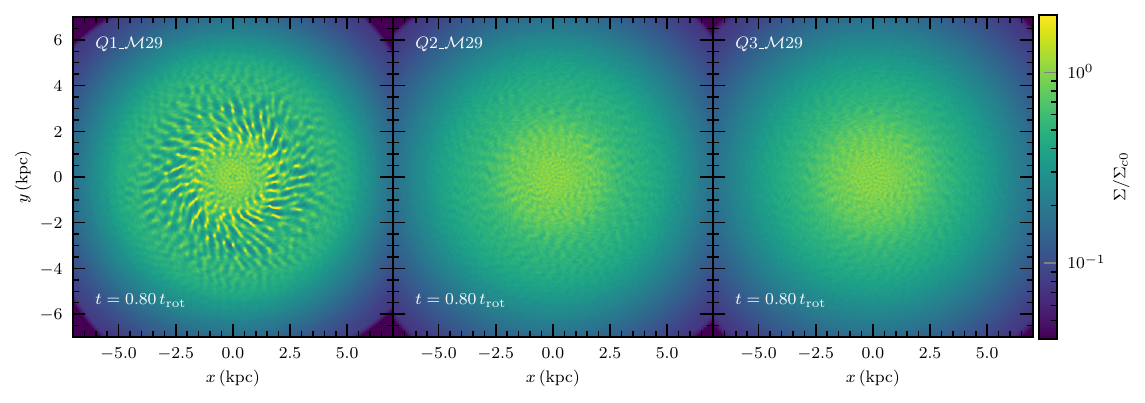}
    \caption{Same as \autoref{fig:projection_Q1_mach29_evolution}, but for runs with $\langle Q \rangle =1,\,2,\,3$ (from left to right) at $t=0.8\,t_{\rm rot}$. We see that the filamentary structures only emerge for the $\langle Q \rangle = 1$ run.}
    \label{fig:projection_toomre_variation}
\end{figure*}

We next examine the dependence of feather formation on the initial $\langle Q\rangle$. Similar to \autoref{fig:projection_Q1_mach29_evolution}, we show the $z$ projections of disc density in \autoref{fig:projection_toomre_variation} for the $\langle Q\rangle =1,\,2,\,3$ (from left to right) runs at fixed $\machc=29$ at identical times $t=0.80\,\trot$. We can immediately see the stark difference between the run with $\langle Q\rangle =1$ and the ones that have a higher $\langle Q\rangle$. For $\langle Q\rangle \in \{2, 3\}$, we see diffuse and faint structures in the disc. This is in contrast to the dense network of filaments that emerge in the $\langle Q\rangle =1$ simulation. We do see development of dense features in the $\langle Q\rangle =2$ run at times later than that shown in the Figure, but the rate of growth of this run is an order of magnitude slower than the $\langle Q\rangle =1$ run, and thus it does not reach significant density contrasts until much later than the other runs. We discuss this case further in \autoref{appendix:Q2_run}, and for the remainder of the main body of this paper we exclusively focus on the $\langle Q\rangle = 1$ runs that form dense structures within a single rotation.

\subsubsection{Dependence on $\machc$} \label{subsubsect: morphology_dependence_machc}

\begin{figure*}
    \centering
     \includegraphics[width=0.99\linewidth]{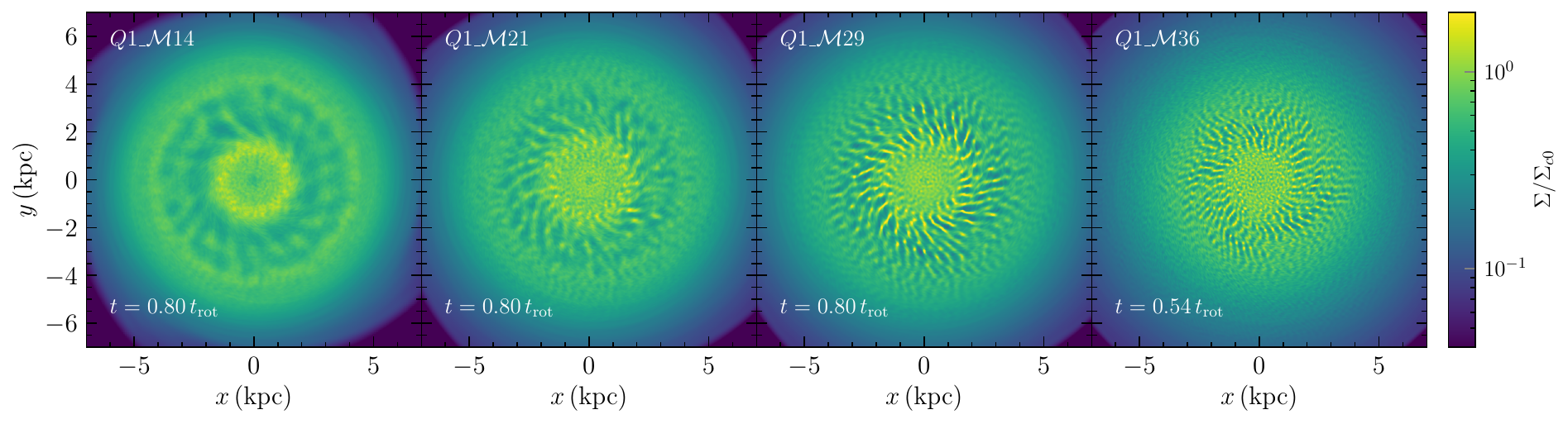}
    \caption{Same as \autoref{fig:projection_toomre_variation}, but for runs with $\machc=14,\,21,\,29,\,36$ (from left to right) and fixed $\langle Q \rangle =1$. All the discs are unstable to filament formation, but filament morphology differs starkly between the different runs, with higher $\machc$ producing more and thinner filaments. The filaments also tend to be shorter at higher $\machc$.}
    \label{fig:projection_Q1_mach_c_variation}
\end{figure*}

We now examine the runs with $\langle Q \rangle = 1$ but varying $\machc$. \autoref{fig:projection_Q1_mach_c_variation} is similar to \autoref{fig:projection_toomre_variation} in that it shows the $z$ projections of the disc density for different runs, but in this case with a common value of $\langle Q \rangle = 1$ but varying $\machc \in \{14, 21, 29, 36\}$. All runs shown are at time $t=0.8\,\trot$ except for the $\mach36$ run, which we show at $t=0.54\,\trot$ because it evolves faster than the other cases and collapses beyond our ability to resolve by $0.8\,\trot$. We see that all the runs form filaments, but that there is a systematic and striking change of feather morphology between the the panels. With increasing $\machc$, the feathers are thinner and more closely-spaced. The $\machc=29$ and $36$ cases also exhibit formation of dense clumps, in contrast to the $\machc=14$ and $21$ runs where we see uniform larger-scale radially extended filaments.

The filaments we see in \autoref{fig:projection_Q1_mach_c_variation} are comparable to the dense filamentary network observed in nearby spiral galaxies with JWST/MIRI observations \citep{thilker_filamentary_NGC628_JWST_2023, meidt_rosolowsky_23} and the lattice or elongated extinction features reported in HST observations \citep{vigne_2006}. In addition to the lattice-like morphology, where the filaments are present throughout the azimuth of the galaxy, the $\machc=36$ case has the filaments arranged in the pattern of spiral arms. This arrangement of filaments is similar to ``spurs'' \citep[see][]{vigne_2006} that are filamentary features connected to spiral lanes. A notable feature of spurs is that they emanate from spiral arms whose pitch angles are lower compared to the pitch angle of spurs. This is also reproduced in the $\machc=36$ case. We defer a more quantitative comparison to observations to \autoref{subsec:comparison_with_observations}.

\subsection{Feather formation timescales} \label{subsect:featherFormation_timescales}
\begin{figure*}
    \centering
    \includegraphics[width=0.99\linewidth]{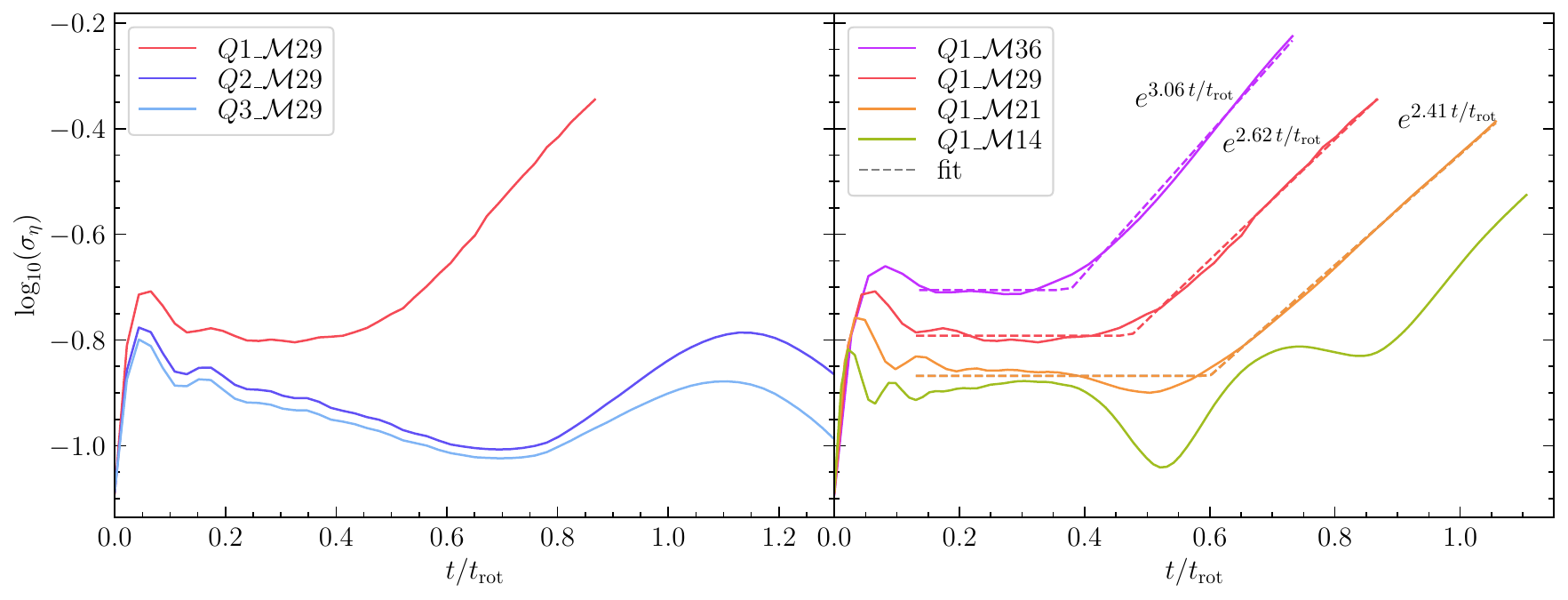}
    \caption{Time evolution of the variance in the logarithmic surface density, $\sigma_\eta$, where $\eta$ is defined in \autoref{eqn:clumping_factor}. The panel on the left is for runs with varying $\langle Q\rangle$ and fixed $\machc$ and the panel on the right for runs with varying $\machc$ but fixed $\langle Q\rangle$, as indicated in the legends. The solid lines represent the simulation data and the dotted lines are piecewise fits with a constant and an exponential part. We see that the $\langle Q\rangle >1$ runs are stable, while there is exponential growth in all the runs with $\langle Q\rangle=1$. For the unstable runs, we see that growth is faster for higher $\machc$. We also observe a decrease in the time of the onset of the instability with increasing $\machc$. We do not provide an analytical fit to the $\machc=14$ case because of its complex behaviour. }
    \label{fig:clumpingFactor_timeEvol}
\end{figure*}

After establishing that the galaxies with $\langle Q \rangle = 1$ forms feathers, we now quantify the timescales of their formation. For this purpose we define the logarithmic surface density,
\begin{equation} \label{eqn:clumping_factor}
    \eta = \ln \left (\frac{\Sigma}{\langle \Sigma \rangle} \right ),
\end{equation}
where $\langle \Sigma \rangle$ is the average surface density of some region of interest. We take our region of interest for this purpose to be an annulus $1\, \si{kpc}$ thick centred at $R = 3\, \si{kpc}$, which is the region where feathers appear (as seen in \autoref{fig:projection_Q1_mach_c_variation}). We use the logarithmic surface density in order to give equal weight to the under- and over-dense regions that differ by a similar factor relative to the average. The variance in the value of $\eta$ over the region of interest, which we denote $\sigma_\eta$, therefore describes the strength of density variations in that region. We show the time evolution of this quantity in \autoref{fig:clumpingFactor_timeEvol}. The left panel is for runs with varying $\langle Q \rangle $ and constant $\machc=29$, while the right panel is for the group with $\langle Q \rangle =1$ but varying $\machc$. The solid lines in both the panels depict the data from the simulations and the dashed lines in the second panel are empirical fits to the data that we describe below.

Starting with the left panel in \autoref{fig:clumpingFactor_timeEvol} which shows runs with varying $\langle Q \rangle$, we see that the variation in the clumping factor rises initially due to the turbulent velocity field causing fluctuations in the density field, and that this rise is similar in all runs. After this initial rise the value of $\sigma_\eta$ stabilises, but at this point the runs begin to diverge. In the $\langle Q \rangle = 1$ run the variations begin to rise again after a short stagnation period, marking the onset of instability. By contrast the runs with $\langle Q \rangle =2,\,3$ show gradual decline followed by a phase of slow growth after around a rotation period -- see \autoref{appendix:Q2_run} for further discussion of the long-time behaviour of the $\langle Q \rangle = 2$ run. For the group of runs with $\langle Q \rangle =1$ but different $\machc$ (right panel), we see an evolution similar to the $Q1\_\mach29$ case for all but one of the runs. That is, an initial jump, followed by a period during which $\sigma_{\eta}$ remains roughly constant, and then an epoch of exponential growth. The $\machc=14$ differs slightly from this pattern, showing a dip in the midst of its steady period before recovering and then embarking on exponential growth. 

While the runs show similar qualitative behaviour, there are quantitative differences. First, we see that the disc develops stronger density fluctuations (higher $\sigma_\eta$) with increasing $\machc$, despite the initial velocity perturbations being unchanged. Second, the exponential growth phase of the instability is systematically earlier for increasing $\machc$. Third, the slopes of the rising curves are higher with increasing $\machc$. We quantify these trends by carrying out a simple $\chi^2$ fit for the time-evolution of $\sigma_\eta$ at times $t\geq 0.15\,\trot$, which is approximately the time of the initial relaxation phase, to an empirically-motivated piecewise function of the form 
\begin{equation} \label{eqn: piece-wise_fit}
    \sigma_\eta(t) = \begin{cases}
     A e^{\omega t_{\rm onset}} & t\leq t_{\rm onset}, \\
    A e^{\omega t} & t > t_{\rm onset},
    \end{cases}
\end{equation}
where $A$ sets the amplitude of $\sigma_{\eta}$ during the steady phase and $\omega$ and $t_{\rm onset}$ quantify the growth rate and the time of the onset of the instability, respectively. We report the best-fitting parameters and their uncertainties in \autoref{table:piece-wise fit} and plot the fits as the dashed lines in the right-hand panel of \autoref{fig:clumpingFactor_timeEvol}; the figure demonstrates that these functional forms fit the data very well, particularly in the time after $t_\mathrm{onset}$. We do not fit for the $\machc=14$ run since it shows a more complex behaviour compared to the other three runs.

For our best fits, we find $t_{\rm onset}\propto \machc^{-1}$ to within $5\%$. Compared to this dependence of the onset time, we find that the growth rate only weakly depends on $\machc$, with the highest growth rate $\omega=3.06 \pm 0.08 \,\trot^{-1}=0.49 \pm 0.01\,\Omega$ for the $\machc=36$ case and the slowest growth rate $\omega=2.41 \pm 0.04 \,\trot^{-1}=0.38 \pm 0.01 \,\Omega$ for the $\machc=21$ run. A decrease in the growth rate is expected as the galaxy gets thicker since the gravitational potential is smeared out in the thick disc case when compared to the thin disc case of equal $Q$ value \citep{meidt_molecular_2022}. To check the robustness of this result against resolution, we carry out a resolution study on the $Q1\_\mach29$ run, which we discuss in \autoref{appendix:resolutionStudy}. The analysis there shows that the run is converged with respect to both the onset time of the instability and the growth rate.

\begin{figure*}
    \centering
     \includegraphics[width=0.99\linewidth]{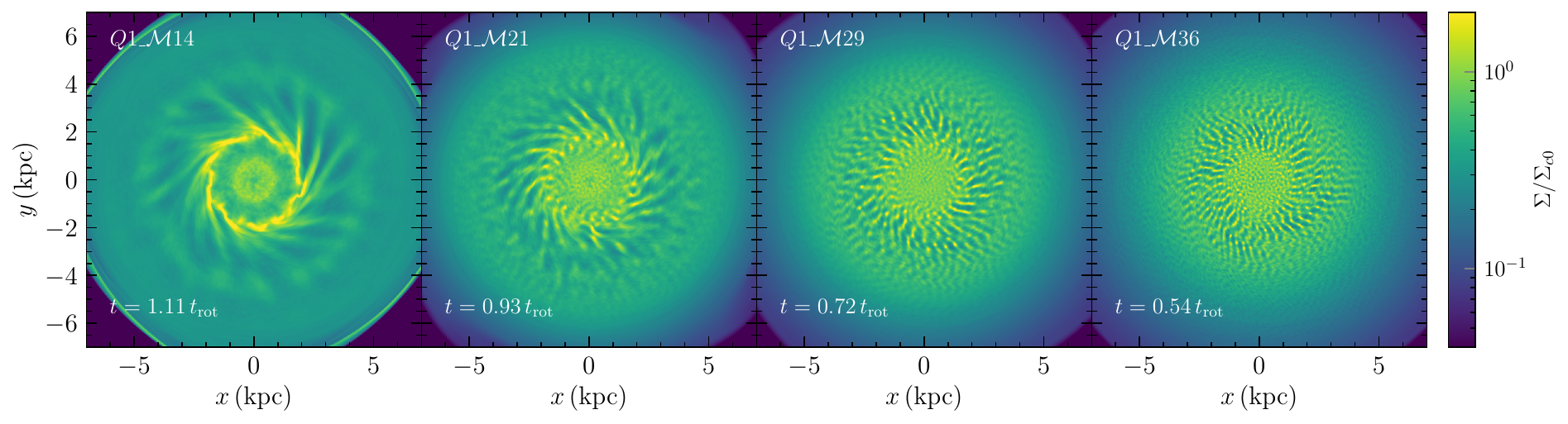}
    \caption{Same as \autoref{fig:projection_Q1_mach_c_variation}, but now showing the different runs at different times. The time is chosen such that the standard deviation of the logarithmic surface density $\log_{10} \sigma_{\eta}= -0.3$ in all runs. We see systematic differences in the filament morphology even when the filament forming regions have similar density contrasts. The filaments are large scale and uniform in the $\machc=14$ and 21 runs compared to the more complex and finely spaced in the $\machc=29$ and 36 cases.}
    \label{fig:projection_Q1_mach_c_variation_sigma_0p3}
\end{figure*}

Since the growth rate of density contrast depends on  $\machc$, it is natural to ask whether the difference in morphology visible in \autoref{fig:projection_Q1_mach_c_variation} is solely the result of the instability being at different evolutionary stages. To answer this question in \autoref{fig:projection_Q1_mach_c_variation_sigma_0p3} we again show projections of all the $\langle Q \rangle =1$ runs, but now choosing times when the different runs are at a common value of $\log_{10}\sigma_{\eta} = -0.3$. The figure shows that, even though we are now examining comparable amplitudes of the instability, substantial morphological differences remain between the runs with differing $\machc$. We quantify these morphological differences in the next subsection. 

\begin{table}
\setlength{\tabcolsep}{3.5pt} 
\caption{Fit parameters for the evolution of the variance in the logarithmic surface density $\sigma_\eta$ (\autoref{eqn:clumping_factor}).} \label{table:piece-wise fit}
\centering
\begin{tabular}{c c c c} 
\hline \hline
Model Name & $A$               & $\omega \trot$        & $t_{\rm onset}/t_{\rm rot}$   \\
\hline
$Q1\_\mach21$   & $0.032 \pm 0.001$ & $2.41 \pm 0.05$ & $0.60 \pm 0.01$ \\
$Q1\_\mach29$    & $0.047 \pm 0.001$ & $2.62 \pm 0.08$ & $0.47 \pm 0.01$ \\
$Q1\_\mach36$    & $0.062 \pm 0.002$ & $3.06 \pm 0.08$ & $0.38 \pm 0.01$ \\
\hline
\end{tabular}
\tablefoot{The fit is performed using a functional form given by \autoref{eqn: piece-wise_fit}. We show the best fits as dashed lines in the right-hand panel of \autoref{fig:clumpingFactor_timeEvol}.}
\end{table}


\subsection{Feather geometry} \label{subsect:featherProperties}

In order to better visualise the morphology of the filamentary structures formed in our galaxies, we project the galaxy in the $(\ln R, \theta)$ plane, where $R$ is the galactocentric radius in $\si{kpc}$ and $\theta$ is the azimuth. This allows easy identification of spiral features, since spirals of the form $R = e^{\theta \tan \alpha}$ appear as straight lines of slope $\tan\alpha$ in the $(\ln R, \theta)$ plane. It also makes the azimuthal periodicity of the features more apparent to the eye. We show this projection in \autoref{fig:projection_Q1_mach_c_variation_sigma_0p3} for the same runs and times as in \autoref{fig:projection_Q1_mach_c_variation}. Since our galaxy rotates  clockwise, the positively-sloped features visible in the plot correspond to trailing spiral features (tips pointing opposite to galactic rotation). We can immediately see that the feathers in the $\machc=14$ and 21 runs appear as straight lines of nearly constant slope with a periodicity in the azimuthal direction. The complex morphology of the  $\machc=29$ and 36 runs appear as thinner spirals with a larger variation in their orientations and lengths compared to the $\machc=14$ and 21 cases. In these runs we also see distinctive ``spur''-like arrangements of the feathers located periodically along spirals, which are visible for example in the distinctive alternating pattern from $\theta\approx -1$ to 0 in the $\machc=36$ run. 

\begin{figure*}
    \centering
    \includegraphics[width = 0.99\linewidth]{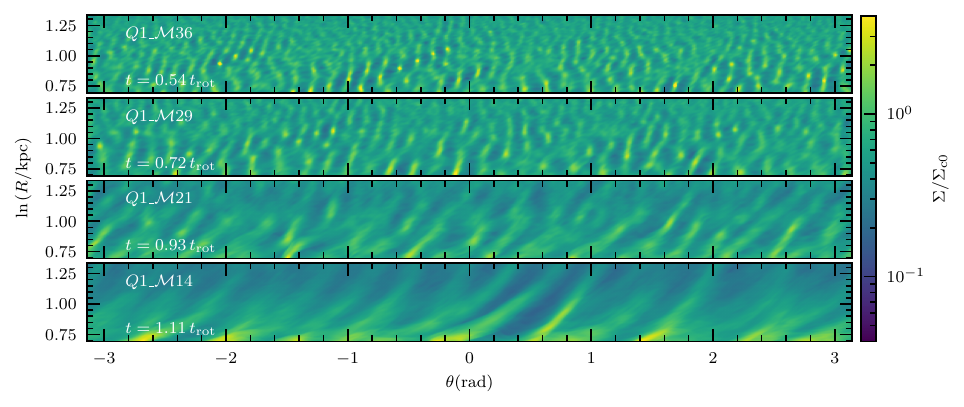}
    \caption{Normalised surface density $\Sigma/\Sigma_{c0}$ in the  $(\ln R, \theta)$ plane for the same runs and times as shown in \autoref{fig:projection_Q1_mach_c_variation_sigma_0p3}. The radial range in this figure is chosen to highlight the annulus from $2-4$ kpc where feather formation is strongest. We can see the feathers appear as sloped lines that are periodic in azimuth, with length, thickness, orientation and spacing that vary with $\machc$. The associated time evolution movies are available online.}
    \label{fig:lnr_theta_plot_1by4}
\end{figure*}

\begin{figure*}
    \centering
    \includegraphics[width = 0.99\linewidth]{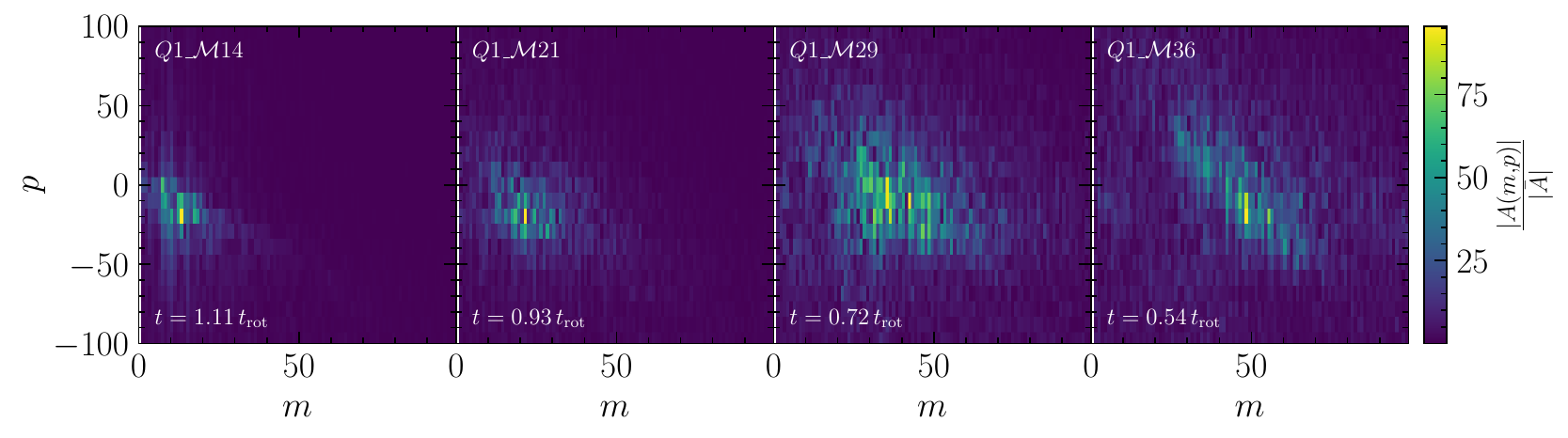}
    \caption{Normalised amplitude of the 2D Fourier transform of the galactic annuli shown in \autoref{fig:lnr_theta_plot_1by4}. Panels show runs with varying $\machc$ (increasing from left to right). We can see that as $\machc$ increases the maximum shifts to higher $m$ and the width of the distribution increases.}
    \label{fig:twoDimenional_fourier_transform}
\end{figure*}

To quantify these morphological differences, we take ``tilted'' 2D Fourier transforms \citep{savchenko_2012, puerari_2014} of the windows shown in \autoref{fig:lnr_theta_plot_1by4}. We define 
\begin{equation}\label{eqn:2d_fourier_transform}
    A(m,p) = \int ^{u_{\rm max}} _{u_{\rm min}} \int ^{2\pi} _{0} \eta (\theta, u) e^{i(m\theta + pu)} \, d\theta\, du,
\end{equation}
where $u = \ln R$, $\eta$ is the logarithmic surface density (defined in ~\autoref{eqn:clumping_factor}), and $(u_{\rm min}, u_{\rm max})$ denote the radial bounds of our annuli. The term $e^{i(m\theta + pu)}$ can be viewed as a filter function that picks up $m$-armed logarithmic spirals with a pitch angle of $\tan \alpha = -m/p$.  We show the amplitude $|A(m,p)|$ normalised to the mean value $\Bar{A} (m,p) = \iint A(m,p) \, dm \, dp/ \iint\, dm\, dp$ in \autoref{fig:twoDimenional_fourier_transform}, plotted with $m$ on the abscissa and $p$ on the ordinate. Different panels show runs with varying $\machc$, increasing from left to right. We see that there is maximum in all the runs, and from the position of this maximum we can read off the preferred azimuthal scale and orientation of the filaments. Shifts in the position of the maximum and the shape of the distribution around it reflect the morphological differences seen between the feathers of different runs. We see that as $\machc$ increases, power shifts to higher $m$ and more positive $p$, and that the distribution in the $(m,p)$ plane broadens. These shifts mirror the more tightly packed and morphologically-complex filamentary pattern we see in going from low to high $\machc$ in \autoref{fig:projection_Q1_mach_c_variation_sigma_0p3}.

\subsubsection{Pitch angle} \label{subsubsect:feather_pitchAngle}

\begin{figure*}
    \includegraphics[width=\linewidth]{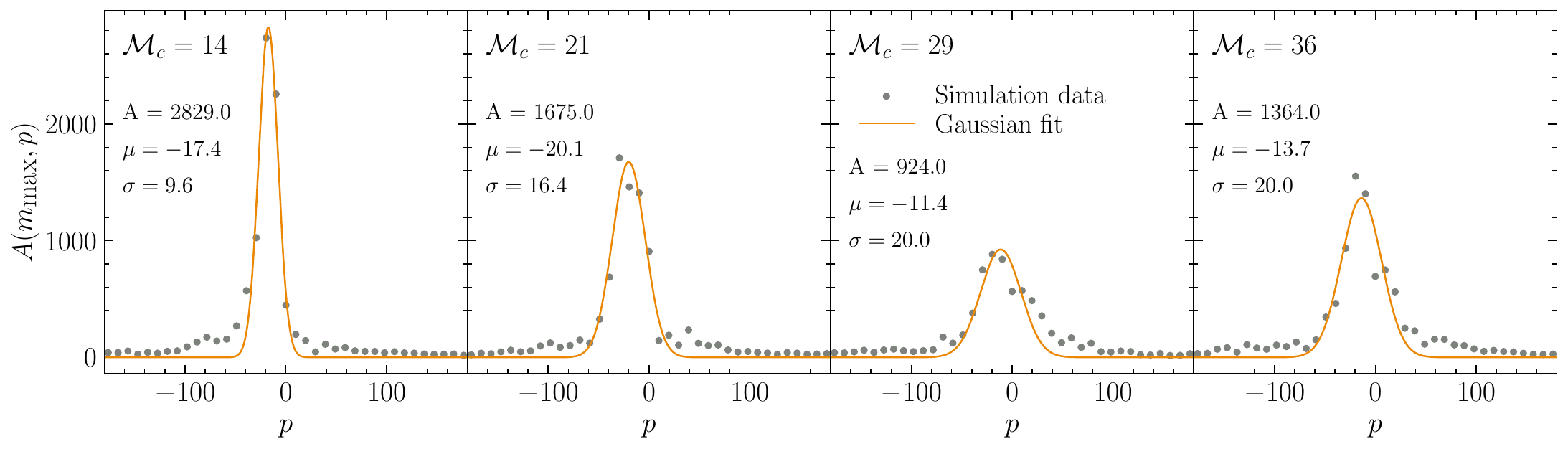}
    \caption{Function $A(m_{\rm max},p)$ for runs with $\langle Q\rangle = 1$ and varying $\machc$ (as indicated in the panels), along with their Gaussian fits. Black points show the amplitude of the Fourier transform $A(m_{\rm max},p)$ as a function of $p$ (c.f., \autoref{eqn:2d_fourier_transform}), where $m_\mathrm{max}$ is the value of $m$ for which $A(m,p)$ has its maximum. Orange lines show Gaussian fits to the data, with the amplitude $A$, mean $\mu$, and dispersion $\sigma$ of each fit as indicated in the legend.}
    \label{fig:a_mMax_vs_p_Gaussian}
\end{figure*}

\begin{figure}
    \centering
    \includegraphics[width=0.99\linewidth]{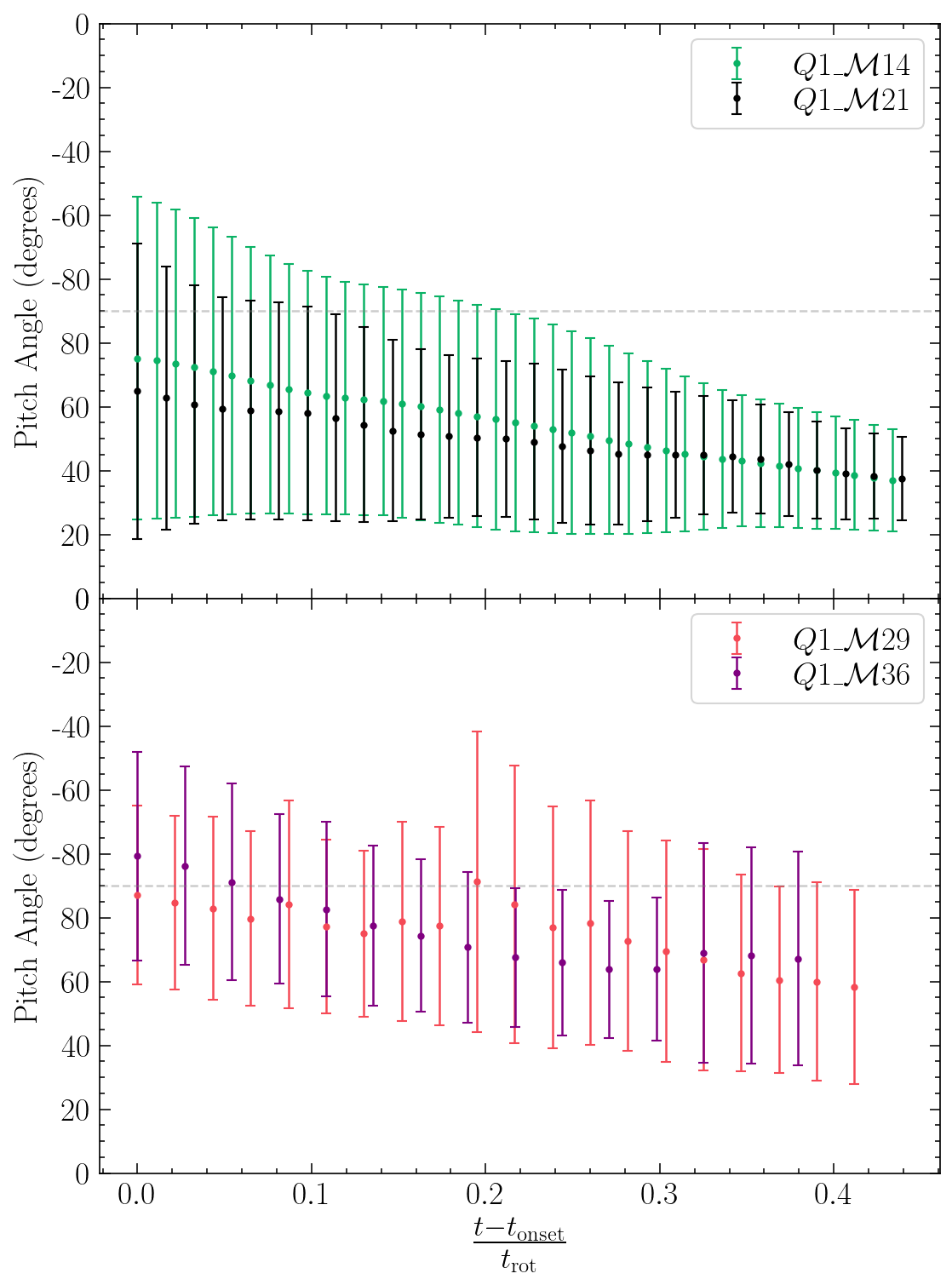}
    \caption{Time evolution of the pitch angle, $\alpha = \tan ^{-1}(-m/p)$, of the filaments in the runs with $\langle Q \rangle =1$ and varying $\machc$. The upper panel shows the $\machc=14$ and 21 runs, and the lower shows $\machc=29$ and 36. All panels show results after the onset of the instability, and all times are normalised by the rotation period of the galaxy at $R=3$~kpc. Circular points represent the mean pitch angle and error bars show the standard deviation. The dashed line shows $\alpha = 90^{\circ}$, that marks the transition between the trailing spirals with a positive pitch angles to leading spiral with negative ones. We see that all the runs show a decreasing trend in the pitch angle as the feathers get sheared due to differential rotation of the galaxy.}
    \label{fig:pitchAngle_with_mach_c}
\end{figure}
The 2D Fourier transform contains information on both the arm number and the pitch angle of the perturbations. To extract the latter we follow the approach of \citet{savchenko_2012} by finding the $m$-mode of maximum power and then examining the range of $p$ value, recalling that the pitch angle $\alpha = \tan^{-1}(-m/p)$. Quantitatively, for each simulation we define $m_\mathrm{max}$ as the value of $m$ at which the maximum of $A(m,p)$ occurs. We then fit $A(m_\mathrm{max},p)$ with a Gaussian functional form in $p$, which provides a reasonably good description of the data. We demonstrate this in \autoref{fig:a_mMax_vs_p_Gaussian}, which shows $A(m_\mathrm{max},p)$ versus $p$, and our best Gaussian fits to it, for the same four snapshots shown in \autoref{fig:lnr_theta_plot_1by4}.

We carry out this procedure for every snapshot of runs $Q1\_\mach14$ to $Q1\_\mach36$, extracting the amplitude $A$, mean $\mu$, and dispersion $\sigma$ of the best-fitting Gaussian at each time. We then take $\alpha = \tan^{-1}(-m_\mathrm{max}/\mu)$ as our best estimate for the pitch angle. For the uncertainty in $\alpha$, we take the dispersion of our fit as an estimate for the uncertainty in $p$, that is $\Delta p = \sigma$, and then propagate it to the uncertainty in pitch angle via standard error propagation, which gives
\begin{equation}
    \Delta \alpha = \left|\frac{d(\tan^{-1} (-m_{\rm max}/p))}{d(-m_{\rm max}/p)} \Delta (-m_{\rm max}/p)\right| = \left|\cos^2 \alpha    \frac{m_{\rm max}}{\mu^{2}} \sigma \right|. 
\end{equation}
We show the pitch angle as a function of time after the onset of the instability that we derive from this procedure in \autoref{fig:pitchAngle_with_mach_c}. The upper panel is for the $\machc = 14$ and $21$ runs and the lower panel for the $\machc=29$ and $36$ cases. We see that, independent of $\machc$, the average pitch angle of all the runs decreases with time. This is also seen visually in the time evolution movies of the surface densities in the $(\ln R, \theta)$ plane (available online). This is expected for material arms being sheared by the differential rotation of the galaxy, since more rapid rotation of the inner regions winds up the arms. In terms of \autoref{fig:lnr_theta_plot_1by4}, this corresponds to a systematic decrease in the slopes of the spiral features with time. This decrease is much more uniform in the $\machc=14$ and $21$ cases when compared to the other two. Moreover, the spread in pitch angle systematically decreases with time for the $\machc=14$ and $21$ runs but remains nearly constant, and a factor of two larger, for the $\machc=29$ and $36$ runs. This difference reflects the more complex and clumpy morphology visible in the higher $\machc$ runs. 

\subsubsection{Spacing} \label{subsubsubsect:feather_spacings}

\begin{table}
\caption{Fit parameters for the lognormal-like fit (see \autoref{eqn:a_m_logNormal_like}) on $A(m)$ from our simulations at the times shown in \autoref{fig:a_m_vs_m_logNormal}. } \label{table:a_m_fit}
\begin{tabular}{llll} 
\hline 
\hline
Model Name & $A_{m}/10^{6}$               & $\mu_{m}$        & $\sigma_m$   \\
\hline
$Q1\_\mach14$   & $2.95 \pm 0.10$ &  $2.72 \pm 0.03 $ & $0.78 \pm 0.03$  \\
$Q1\_\mach21$   & $3.88 \pm 0.10$ &  $3.40 \pm 0.03$ & $0.80 \pm 0.02$  \\
$Q1\_\mach29$    & $5.59 \pm 0.15$ & $3.81 \pm 0.02$ &$0.69 \pm 0.02$  \\
$Q1\_\mach36$    & $7.87 \pm 0.23$& $4.11 \pm 0.03$ & $0.82 \pm 0.02$ \\
\hline
\end{tabular} 
\end{table}

\begin{figure*}
    \includegraphics[width=\linewidth]{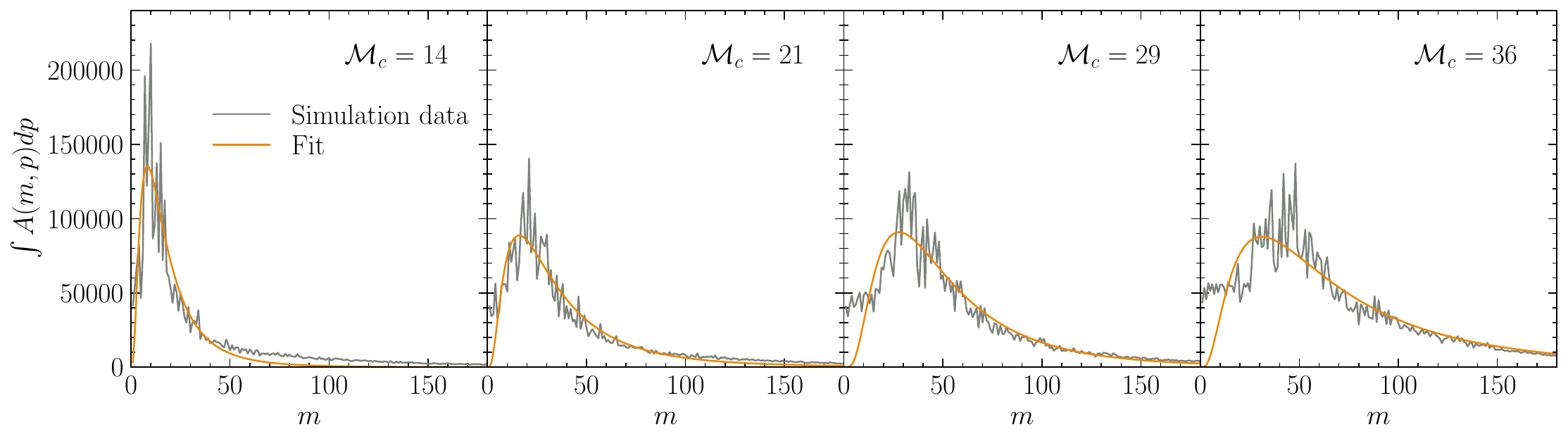}
    \caption{Amplitude of the 2D Fourier transform $A(m,p)$ of the logarithmic surface density (see \autoref{eqn:2d_fourier_transform}) in the region $2$-$4\, \si{kpc}$, integrated over $p$. Different panels are for runs with $\langle Q \rangle=1$ and varying $\machc$, evaluated at the same times as in \autoref{fig:projection_Q1_mach_c_variation_sigma_0p3}. The grey curves are simulation data and the orange ones are the empirical fits to the data (see text for details). We can see how the power shifts toward higher values of $m$ with increasing $\machc$.}
    \label{fig:a_m_vs_m_logNormal}
\end{figure*}

To quantify feather spacing we integrate the 2D transform over the $p$ direction to yield $A(m) = \int A(m,p) \, dp$. We find that $A(m)$ does not vary appreciably after $t_{\rm onset}$, and thus for simplicity we carry out our analysis only on the snapshots shown in \autoref{fig:lnr_theta_plot_1by4}; results for other snapshots are qualitatively identical. As with our analysis of the pitch angle, we fit an empirically chosen functional form to $A(m)$; since $m$ is a positive-definite quantity, we use a lognormal-like rather than a normal form, 
\begin{equation} \label{eqn:a_m_logNormal_like}
    A(m) = \frac{A_m}{m\sigma_{m} \sqrt{2\pi}} \exp \left [ \frac{(\log m - \mu_{m})^{2}}{2\sigma_{m}^{2}} \right ],
\end{equation}
where $A_m$ is the amplitude, $\mu_m$ is the location of the peak in $\log m$, and $\sigma_m$ is the width in $\log m$. In \autoref{fig:a_m_vs_m_logNormal} we show $A(m)$ along with our best empirical fits, the parameters for which we report in \autoref{table:a_m_fit}. The figure demonstrates that our chosen functional form provides a reasonable description of the data. From the empirical fit we extract the median, given by $m_{\textrm{fil}} = \exp {(\mu_{m})}$, as a representation of the dominant azimuthal mode, and the distribution's $16$th and $84$th percentiles given by $\exp{ (\mu_m \mp \sigma_{m} )}$, as its lower and upper spread. 

\begin{figure}
    \centering
    \includegraphics[width=0.99\linewidth]{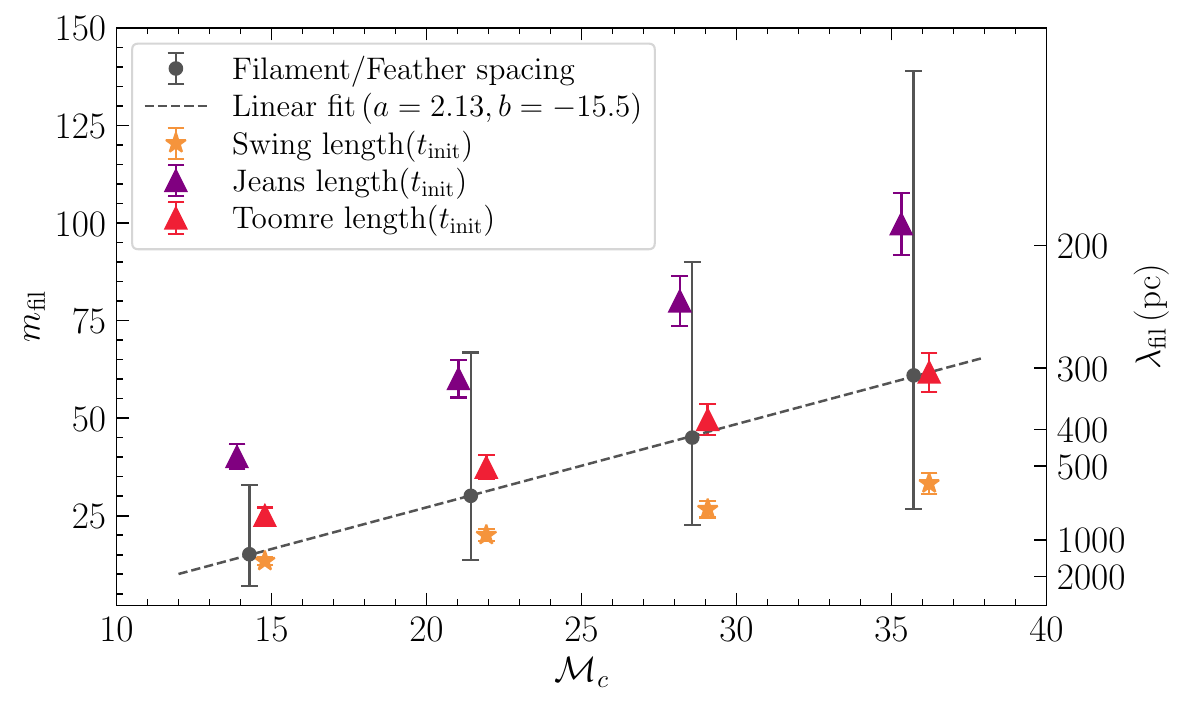}
    \caption{Azimuthal mode of the filaments ($m_{\rm fil}$) and various instabilities, shown with the $\machc$ $(\sim \vc/\sigma)$ of our galaxies. Black circular points with error bars show the mean and 16th to 84th percentile range of the $m_{\rm fil}$ and corresponding physical spacing $\lambda_{\rm fil}$ for the snapshots shown in \autoref{fig:lnr_theta_plot_1by4}. The dashed line is a linear fit to the median values of $m_{\rm fil}$ (\autoref{eqn:m_linear_fit}). Purple, red, and orange points with error bars represent the radial averages $\langle\lambda\rangle$ and standard deviations $\sigma_\lambda$ (\autoref{eq:lambda_mean} and \autoref{eq:lambda_stddev}) of the wavelength of the most unstable Toomre mode, the Jeans length and the Swing amplification length -- see main text for details. These points are slightly offset from their overlapping $\machc$ values for visibility. The simulations show an increase in the median value of $m_{\rm fil}$ and a broadening of the 16th to 84th percentile range with increasing $\machc$. The median filament spacing agrees well with the Toomre length of the $\machc = 21, 29$ and $36$ runs, and with the swing length for $\machc = 14$ run.}
    \label{fig:m_with_mach_c}
\end{figure}

We show $m_{\textrm{fil}}$ as a function of $\machc$ for our four runs in \autoref{fig:m_with_mach_c}. The right vertical axis of this plot shows the physical spacing corresponding to this value of $m_\mathrm{fil}$, which is $\lambda_{\rm fil} = 2\pi R_{\textrm{gal}}/m_{\textrm{fil}}$, where $R_{\textrm{gal}}=3$~kpc is the mean radius of the annulus we use in our analysis. We find that the $m_{\rm fil}$ varies approximately linearly with $\machc$. A $\chi^2$ fit to this relationship of the form
\begin{equation} \label{eqn:m_linear_fit}
    m_{\rm fil} = a\machc + b, 
\end{equation}
yields $a = 2.13 \pm 0.02 $ and $b = -15.5 \pm 0.6$, and we show this fit via the black dashed line in \autoref{fig:m_with_mach_c}.

\subsection{Feather density structure and evolution}

\begin{figure*}
    \centering
    \includegraphics[width=0.99\linewidth]{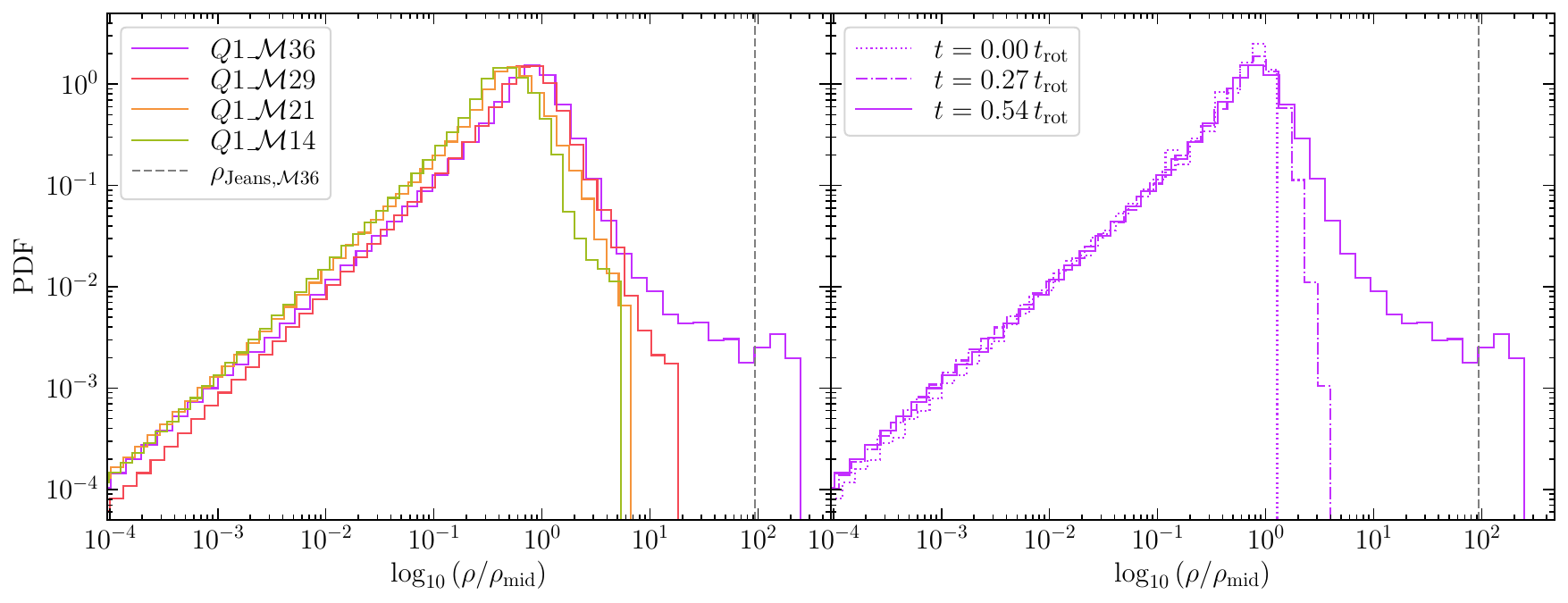}
    \caption{Mass-weighted probability density function (PDF) of the $\log_{10}$ of the density normalised by the radially averaged mid-plane density, for the snapshots shown in \autoref{fig:lnr_theta_plot_1by4}. The grey dashed line depicts the Jeans density (see text for details), at which we expect numerical effects to dominate. The left panel shows all the runs at times similar to \autoref{fig:lnr_theta_plot_1by4}, and the right panel shows the time evolution of the $Q1\_\mach36$ case leading up to the time in the left panel. We see that all the runs in the left panel form high-density tails that represent feathers and that the densest end of the tail is reaching the limit imposed by the resolution of our simulations.
    \label{fig:density_contrasts}
    }
\end{figure*}

Finally, we examine the three-dimensional density structure of the feathers. Our goal here is understanding how this depends on $\machc$, and to see whether the feathers are likely to be sites for star formation.  \autoref{fig:density_contrasts}, where we shows the mass-weighted probability density function (PDF) of $\log_{10}$ of $\rho/\rho_{\rm mid}$, where $\rho$ is the density of the gas and $\rho_{\rm mid}$ is the radial average of the initial density of the region of interest. We do this for the cylindrical region bounded by $1$~kpc thick radial annuli centred at $3$~kpc. The left panel is for all the ``feather forming'' runs with $\langle Q \rangle = 1$ and varying $\machc$ at times when they are similarly evolved; see snapshots in \autoref{fig:lnr_theta_plot_1by4}. In the right panel we show the PDF at several different times for the $Q1\_\mach36$ run.

In this plot, the stratified medium of the background galaxy corresponds to the parts of the PDF with $\rho/\rho_{\rm mid}\leq1$. Filaments and further dense regions inside them are represented in the higher-density tails. Looking at the left panel, we can see that all of our runs form the high-density tail, which is expected from feather-forming galaxies. However, we also see a systematic trend that the higher $\machc$ runs have longer tails extending towards higher densities. This densest part contains little mass and is largely de-coupled from the initial phase of the instability, and is instead driven by self-gravity.

To confirm the importance of self-gravity, in the right panel we add a vertical dashed line marking the Jeans density for the $Q1\_\mach36$ run, defined as $\rho_{\rm Jeans}=(c_{\mathrm{s}}/4\Delta x_{\mathrm{min}})^{2} (\pi/G)$, which is the density at which self-gravitating structures become unresolved on the grid scale \citep[formally, our expression is the density for which the Jeans length is equal to $4\Delta x_\mathrm{min}$]{TrueloveEtAl1997}. Here, $c_{\mathrm{s}}$ is the sound speed and $\Delta x_{\mathrm{min}}$ is the length of the grid cells. We see that the high-density tail of the PDF does not reach a steady-state, and instead rises until it reaches the Jeans density at our resolution. This likely indicates that the high-density tail of the PDF is made up of gas in a state of runaway collapse, where star formation is likely to occur in the future. A detailed study of how this star formation proceeds is beyond the scope of this work.

\section{Discussion} \label{sec:Discussion}

In this section we discuss our results in the light of existing observational, numerical and theoretical works and point out limitations of our approach. In \autoref{subsect:natureOfInstability} we comment on the nature of instability that is responsible for feather formation, and in \autoref{subsec:comparison_with_observations} we compare the filament spacing found in our simulations with observations of nearby galaxies. Lastly, in \autoref{subsect: limitation_futureWork}, we outline the limitations of our simulations and hint on future avenues.

\subsection{Nature of instability} \label{subsect:natureOfInstability}
In order to better understand the physical mechanism responsible for feather formation in our simulations, we compare the mean azimuthal spacing of the filaments calculated in \autoref{subsubsubsect:feather_spacings} to the length scales of various instabilities expected to be present in the disc: the wavelength of the most unstable Toomre mode \citep{toomre_1964}, the Jeans length \citep{Jeans1902} in the galactic mid-plane, and the length scale of the Swing amplification mechanism \citep{fuchs_2001, kim_ostriker_2002, meidt_2024}. These are given by $\lambda_{\rm Toomre} = 2 \cs ^{2}/G\Sigma$,  $\lambda_{\textrm{Jeans}} = (\pi \cs ^{2}/G\rho_{\rm mid})^{1/2}$ and $\lambda_{\rm Swing} = \lambda_{\textrm{Jeans}} X_{J}$, respectively, where $\rho_{\rm mid}$ is the mid-plane density of the galaxy and $X_{J}$  measures the characteristic length of the swing amplification in terms of the Jeans length and has been found to be $\approx 3$ for a wide range of perturbing wavelengths \citep{meidt_2024}; we adopt $X_{J} = 3$ exactly for our comparison. 

These length scales all depend on the local properties of the disc, and thus vary with position and time within our analysis region. In order to reduce this complexity to a single parameter for each run, we take averages over our region of interest using the initial configurations in the simulations. Specifically, for each length scale $\lambda$, we defined
\begin{equation}
    \langle \lambda \rangle  = \frac{1}{R_\mathrm{out}-R_\mathrm{in}} \int ^{R_{\rm out}} _{R_{\rm in}} \lambda(R, t=0)\, dR,
    \label{eq:lambda_mean}
\end{equation}
where $R_{\rm in} = 2$~kpc and $R_{\rm out} = 4$~kpc are the bounding radii of our annulus and $\lambda(R,t=0)$ represents the length scale of interest evaluated for the initial disc state at radius $R$. We also compute the standard deviations of $\lambda$ over this radial range, defined as
\begin{equation}
    \sigma_\lambda = \left\{\frac{1}{R_\mathrm{out}-R_\mathrm{in}} \int ^{R_{\rm out}} _{R_{\rm in}} \left[\lambda(R, t=0)-\left\langle\lambda\right\rangle\right]^2\, dR
    \right\}^{1/2}.
    \label{eq:lambda_stddev}
\end{equation}

We plot $\langle \lambda\rangle$ for each of our candidate length scales -- Jeans, Toomre, and Swing -- on top of values of $\lambda_\mathrm{fil}$ measured from our simulations in \autoref{fig:m_with_mach_c}; in this plot, the errorbars on the points are the standard deviations of the corresponding quantities. Comparing the theoretically-predicted length scales with the measured values of $\lambda_{\rm fil}$, first we note that except for the $\lambda_{\rm Jeans}$ in the $\machc=14$ case, all lie within the spread of the measured filament spacings, but that $\lambda_{\rm Jeans}$ is consistently the farthest from the measured value of $\lambda_{\rm fil}$. The other two length scales are closer to the measurements, with $\lambda_{\rm Toomre}$ lying very close to $\lambda_\mathrm{fil}$ for the $\machc = 29$ and $36$ runs, and $\lambda_{\rm Swing}$ matching it very closely for the $\machc = 14$ case. The $\machc=21$ run lies at a transition for these two behaviours.

An appealing interpretation of this trend is that filament formation is driven by the Toomre instability in the two higher $\machc$ cases, while Swing amplification dominates at lower $\machc$. Such a transition is expected theoretically, because the threshold value of $Q$ required to trigger Toomre instability increases with disc thickness \citep{wang_equilibrium_2010}, and the low $\machc$ discs are thicker than the high $\machc$ ones. Thus for our series of runs at fixed $Q$, we might expect Toomre instability to dominate in high $\machc$ discs that are thin enough to trigger Toomre instability, and for the somewhat slower-growing Swing mechanism to dominate in thicker discs that are Toomre-stable. Our results also appear to be quantitatively consistent with this hypothesis. While the exact $Q$ threshold for Toomre instability in a finite-thickness disc is somewhat sensitive to the exact vertical distribution of the gas, \citet{wang_equilibrium_2010} find typical thresholds of $\simeq 0.7-0.8$ for discs with $\machc$ in the range $16$--$24$. Examining \autoref{fig:toomre_classical_radialProfile}, it is clear that our discs do not quite reach such small $Q$ values in the feather forming region from $2$--$4$~kpc, and thus it seems likely that our $\machc =  14$ and $21$ cases are Toomre stable, consistent with our finding that Swing amplification dominates in these. 

Swing amplification also offers an explanation for the structure formation seen at very late times in the $\langle Q \rangle = 2$ run, which has a $Q$ value well above the threshold for Toomre instability at all radii (\autoref{appendix:Q2_run}).

In observations it can be even more challenging to single out the mechanism that is directly responsible for filament formation since we do not have the option to simply ``switch-off'' physical processes that affect gas flows. Still, there is one observational study that analyses the emergence of galactic filaments as a consequence of gravitational instability \citep{meidt_rosolowsky_23}. They find that their four filament-forming galaxies have $Q\geq1$, that is, above the more conservative Toomre threshold for razor-thin discs. While the galaxies could still form filaments via the Swing amplification, we point out that the inclusion of the major stellar component \citep{leroy_phangs_co_2021} will bring down this $Q$ value potentially making them gravitationally unstable \citep{romeo_falstad_2013}. \citeauthor{meidt_rosolowsky_23} also find that the filament spacing in these galaxies is more comparable to the estimated Jeans length, rather than the Toomre length, which however, will also change with the inclusion of the stellar component. Given the high degree of complexity of the filaments observed in these galaxies, we believe that it is more likely that they are formed due to the Toomre-like galactic scale instability. We compare the filament spacings from observations with our simulations in the next section.

\subsection{Comparison with observations} \label{subsec:comparison_with_observations}
\renewcommand{\arraystretch}{1.3}
\begin{table}[]
\setlength{\tabcolsep}{3.5pt} 
\caption{Physical properties of the galaxies we use for comparison with our simulations.}
\centering
\label{tab:observations_table}
\begin{tabular}{lllll}
\hline 
\hline
Galaxy   & $\lambda_{\rm fil}$  & $v_{\rm c}$ & $\log_{10} \rm SFR$ & $\log_{10}  M_{\rm stellar}$ \\
 & (pc) &  $(\kms)$& $ (\rm M_{ \odot}\, yr^{-1})$ & ~ $\rm (M_{\odot})$  \\
\hline 
NGC 628  & $388^{+49} _{-102} ~(1)$   & $144.8^{+108.0}_{-37.3}$       & 0.24                                   & 10.34                                   \\
NGC 1365 & $1051^{+234} _{-404}~(1)$  &$186.7^{+26.8}_{-18.1}$     & 1.24                                   & 10.99                                   \\
IC 5332  & $660^{+228} _{-240}~(1)$  & $120.0^{+20.6}_{-20.6}$      & -0.39                                  & 9.67                                    \\
NGC 7496 & $547^{+105} _{-113}~(1)$  & $147.7^{+4.8}_{-5.6}$        & 0.35                                   & 10.00                                   \\
NGC 1300 & $374^{+19}_{-19}~(2)$     & $183.3^{+45.2}_{-21.6}$      & 0.07                                   & 10.62                                   \\
NGC 1566 & $264^{+88}_{-88}~(2)$     & $220.6^{+105.9}_{-54.4}$     & 0.66                                   & 10.79                                   \\
NGC 4254 & $568^{+348}_{-348}~(2)$   & $183.2^{+6.5}_{-4.1}$        & 0.49                                   & 10.42                                   \\
NGC 4548 & $197^{+9}_{-9}~(2)$       & $192.9^{+23.1}_{-14.8}$      & -0.28                                  & 10.70                                   \\
NGC 4579 & $299^{+48} _{-48}~(2)$    & $314.0^{+43.3}_{-28.8}$      & 0.33                                   & 11.15                                   \\
NGC 5055 & $525^{+215}_{-215}~(2)$   & $181.1^{+11.9}_{-11.9}$      & 0.28                                   & 10.72      \\   
\hline
\end{tabular}
\tablefoot{Star formation rates and stellar masses from \cite{leroy_phangs_co_2021} for all galaxies except NGC 5055, which is from \cite{leroy_2019_galex_wise}. Feather spacing ($\lambda_{\rm fil}$) from 1 -- \cite{meidt_rosolowsky_23} 2 -- \cite{vigne_2006} tabulated in \cite{mandowara_physical_2022}; circular velocities ($\vc$) from \cite{lang_phangs_rotation_2020} for all except NGC 5055 and IC~5332, which are from \cite{mcGaugh_schombert_2015} and \cite{meidt_2018}, respectively.}
\end{table}

\begin{figure}
    \centering
    \includegraphics[width=0.99\linewidth]{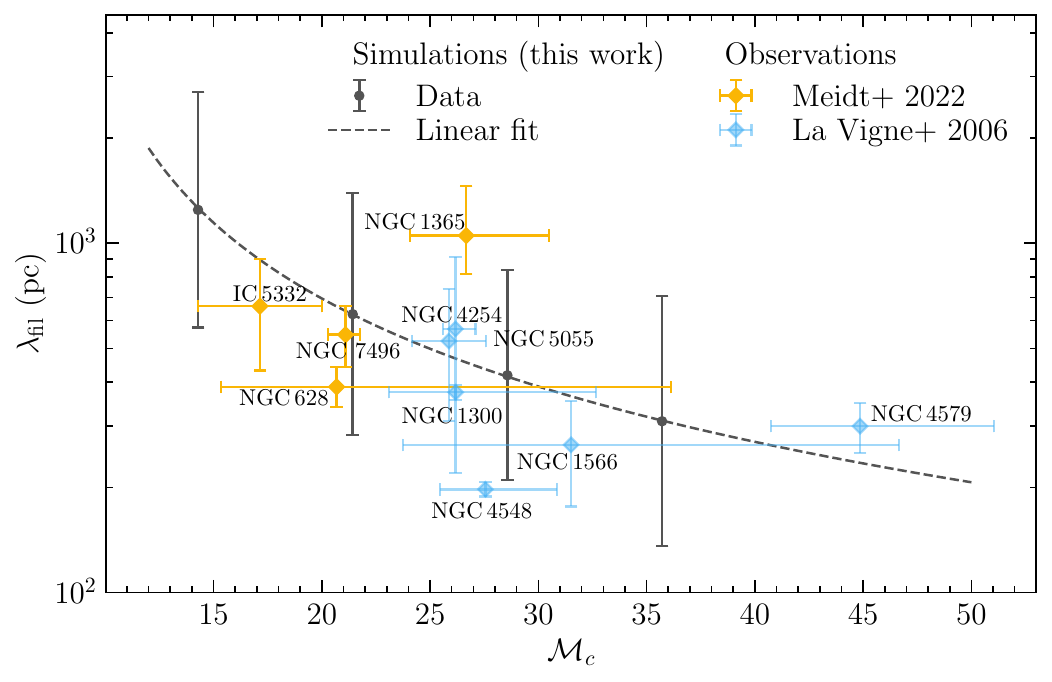}
    \caption{Physical spacing of filaments as a function of the rotational Mach number, $\machc$ ($\sim \vc/\sigma$), in our simulations (black circles; c.f., \autoref{subsubsubsect:feather_spacings}) compared with observations (yellow and blue diamonds). The yellow diamonds are JWST/MIRI observations \citep{meidt_rosolowsky_23} and the blue diamonds are HST observations \citep{vigne_2006}. We see that, except for NGC~$1365$ and NGC~$4548$, the filament spacing extracted from our simulations agrees within the uncertainty ranges of the ones reported in the observations.}
    \label{fig:spacings_comparison_obs_linearTheory}
\end{figure}

We now compare the filament spacing extracted from our simulations with those observed in nearby galaxies, adopting a fiducial radius of $3$~kpc. We plot the filament spacing in our simulations together with observed filament spacings as a function of $\machc$ in  \autoref{fig:spacings_comparison_obs_linearTheory}. In this plot we show the measured filament spacings from our simulations as black points with error bars, and the values shown match the right-axis values indicated in \autoref{fig:m_with_mach_c}; similarly, the dashed line shows the empirical curve given by \autoref{eqn:m_linear_fit}, and is identical to the dashed black line in \autoref{fig:m_with_mach_c}. 

We draw the observations (shown as diamonds in \autoref{fig:spacings_comparison_obs_linearTheory}) from two sources: the JWST/MIRI $11.3\,\mu$m observations reported by \citet[their Table 1]{meidt_rosolowsky_23}, which we plot in yellow, and the HST archival data of \citep[shown in blue]{vigne_2006}, and with the feather spacing of these sources taken from the tabulation of \citet[their Table 4]{mandowara_physical_2022}. \citet{thilker_filamentary_NGC628_JWST_2023} show that there is good agreement between filament catalogues derived from optical and IR data, and thus for the analysis that follows we will not differentiate between the two data sources. We extract the filament spacings and values of $\machc$ that we plot as follows. For the JWST observations, our central points correspond to the reported $50$th percentile of the filament spacing, and our lower and upper error bars to the $16$th and $84$th percentiles, respectively, for the region $R\leq 4 \, \si{kpc}$; however, the choice of radius has little effect since \citeauthor{meidt_rosolowsky_23} find no appreciable variation in filament spacing with galacto-centric radius. For the HST observations, \citeauthor{mandowara_physical_2022} report filament spacings at various galacto-centric radii, so to estimate the spacing at $R=3$~kpc we assume that each galaxy is characterised by a single, global $m_{\rm fil}$ at all radii, which we extract by carrying out a simple $\chi^2$ fit for $m_\mathrm{fil}$ to the radius-dependent values of $\lambda_\mathrm{fil}$ reported in \citeauthor{mandowara_physical_2022}'s Table 4; we then compute $\lambda_{\rm fil}= 2\pi R/ m_{\rm fil}$ at $R = 3$~kpc. For $\machc$ in both sets of observations, we assume a fixed $c_\mathrm{s} = 7$ km s$^{-1}$ for all galaxies, characteristic of the warm neutral medium \citep{naomi_2023_hi}, and take our circular velocities $v_\mathrm{c}$ from the saturated circular velocities of all galaxies calculated from CO~($2$-$1$) emission maps taken from \citet[their Table 4]{lang_phangs_rotation_2020}; we derive error bars in $\machc$ from their quoted uncertainties in $v_\mathrm{c}$, assuming no uncertainty in $c_\mathrm{s}$. The only exceptions to this are the galaxies IC 5332 and NGC 5055, which are not included in this sample. For NGC 5055, we use the H$\,\textsc{I}$ rotation curve velocity from \citet{mcGaugh_schombert_2015} and for IC 5332 we adopt $\vc = 120 \pm 20.6\, \kms$, derived from a semi-empirical model that links the rotation curve to the stellar mass \citep{meidt_2018}, where the spread is propagated from the uncertainty in the galaxy's inclination angle \citep{leroy_phangs_co_2021}. These properties, along with the star formation rates (SFR) and total stellar masses ($M_{\rm stellar}$),are tabulated in \autoref{tab:observations_table}. The SFR and the $M_{\rm stellar}$ are taken from \cite{leroy_phangs_co_2021} for all but NGC 5055, which is taken from \cite{leroy_2019_galex_wise}. Our sample consists of late-type main sequence star-forming galaxies.

From \autoref{fig:spacings_comparison_obs_linearTheory}, we can see that the filament spacing from our simulations agrees well with the observations despite the simplicity of our models. All but NGC 1365 fall within a factor of $\sim2$ of the median of our filament spacing. The trend of decreasing filament spacing with increasing $\machc$ predicted by the simulations is reproduced in the observational data, albeit with a fair amount of scatter since the data are sparse at both high and low $\machc$. The comparison is currently limited by the unavailability of kinematic data for many galaxies whose filament spacings have already been tabulated.

\subsection{Limitations and future work} \label{subsect: limitation_futureWork}

Our simulations consist of self-gravitating, isothermal gas. As a result, we miss some key physical processes pervasive in the ISM that are dynamically relevant on galactic scales. These include magnetic fields, cosmic rays, and feedback processes such as supernovae and winds from massive stars. Moreover, we use static analytical potentials to approximate the gravitational interaction of stars and dark matter with the gas. Therefore, we miss the back-reaction of the former with the latter and vice-versa. 

The inclusion of each of these aforementioned processes may affect feather formation and structure, acting as an additional source or damping mechanism for the perturbations or directly impacting the nature of the instability. For example, magnetic fields can have a stabilising effect due to the additional magnetic pressure. However, they are also known to destabilise the disc via Parker instability \citep{parker_1966, bastian_2019, arora_2023} and the magneto-Jeans instability \citep{kim_ostriker_2002}. Further, since we expect star formation in feathers, energy and momentum injection due to feedback from stars may also affect their morphology. Finally, the presence of a live stellar component may destabilise the disc \citep{romeo_wiegert_2011_q_stability}, which might lead to a shift in the $Q$ thresholds for feather formation.

We point out, however, that the omission of these physical processes is intentional and motivated by our goal of conducting a clean numerical experiment to determine the role of gravity, rotation, and pressure alone in the formation of filaments, and to determine their morphological properties. Future studies will aim to incorporate more physical processes in order to study feather formation in more realistic galaxies. Future studies may also investigate the distribution of filament properties such as their masses, their velocity dispersion, and aspect ratios. The relative orientation of the filaments with respect to the local magnetic field is also of interest. These properties could help us gain insight into the variations of the initial conditions of star-forming regions residing inside feathers \citep[for example;][]{thilker_filamentary_NGC628_JWST_2023}. 

\section{Summary} \label{sec:conclusions}
We simulate isothermal, self-gravitating, isolated disc galaxies that are initialised in equilibrium, with the aim of capturing structure formation via global gravitational instability. Our models are intentionally simplified, allowing us to isolate the dependence of structure formation on two dimensionless parameters: the radially averaged Toomre parameter $\langle Q\rangle $ and the rotational Mach number, $\machc$. Our main conclusions are summarised below.

\begin{itemize}
    \item Galaxies with $\langle Q \rangle = 1$ in the region $R\geq 2$~kpc form dense filaments and feathers within a single rotation, while galaxies with $\langle Q \rangle = 2,3$ fail to form them within similar evolution times. This demonstrates that filaments can form solely due to galactic-scale instability driven by self-gravity overcoming the stabilising effect of rotation and thermal pressure. Thus filament formation occurs without the need for an external stellar spiral potential, magnetic fields, or supernova feedback. This does not prove that these other mechanisms cannot alternatively generate filaments, or that they do not modify the properties of filaments formed via gravitational instability; it simply establishes that gravitational instability alone is sufficient.
    \item The dense filaments that form in our simulations are kiloparsec long and semi-regularly spaced in azimuth. Their morphology and their density contrasts vary systematically with $\machc$: morphological complexity increases with $\machc$ and filament spacing varying as roughly $\machc^{-1}$.
    Filament pitch angle decreases with time in all runs due to shearing by differential rotation, but the dispersion in pitch angle is larger at higher $\machc$ as well. Filaments with a higher $\machc$ also show density contrasts with longer tails that extend towards higher densities.
    \item The time required for onset of the filament forming instability varies $\machc^{-1}$ as well; by contrast, the growth rate of variations in the density depends only weakly on $\machc$, and is typically 30-50\% of the galactic angular velocity. Highest for the run with higher $\machc$. 
    \item The filament spacings and their dependence on $\machc$ found in our simulations are in good agreement with the characteristic length scales of the Toomre instability at higher $\machc$, and with the length scale of the Swing amplification mechanism for lower $\machc$. They are also consistent within observational uncertainties with values measured in nearby galaxies.
\end{itemize}

Our results demonstrate that the filamentary ISM structures revealed by modern telescopes and facilities such as the JWST and ALMA in nearby galaxies with unprecedented resolution can at least potentially be understood as arising from very simple physical processes. Our numerical experiments show that self-gravity alone is sufficient to produce structures whose characteristics are in quantitative agreement with the observations. However, we are yet to see how this might change with the inclusion of other mechanisms such as magnetic fields, cosmic rays, or stellar feedback. In a forthcoming companion study we will quantify some of these effects.

\begin{acknowledgements}
R.~A.~acknowledges funding by the Hamburg State Graduate Funding Program (HmbNFG) and heartfully thanks Sharon Meidt, Shivan Khullar, and Chris Matzner for insightful discussions and their enthusiasm. C.~F.~acknowledges funding provided by the Australian Research Council (Discovery Project grants~DP230102280 and~DP250101526), and the Australia-Germany Joint Research Cooperation Scheme (UA-DAAD). M.~R.~K.~acknowledges support from the Australian Research Council through Laureate Fellowship FL220100020. We further acknowledge high-performance computing resources provided by the Leibniz Rechenzentrum and the Gauss Centre for Supercomputing (grants~pr32lo, pr48pi and GCS Large-scale project~10391), the Australian National Computational Infrastructure (grants~ek9 and jh2) and the Pawsey Supercomputing Centre (projects~pawsey0807 and pawsey0810) in the framework of the National Computational Merit Allocation Scheme and the ANU Merit Allocation Scheme. The simulation software, \texttt{FLASH}, was in part developed by the Flash Centre for Computational Science at the University of Chicago and the Department of Physics and Astronomy at the University of Rochester.
\end{acknowledgements}

\bibliographystyle{aa}
\bibliography{arora_federrath_library}

\begin{thebibliography}{62}
\expandafter\ifx\csname natexlab\endcsname\relax\def\natexlab#1{#1}\fi

\bibitem[{{Arora} {et~al.}(2024){Arora}, {Federrath}, {Banerjee}, \& {K{\"o}rtgen}}]{arora_2023}
{Arora}, R., {Federrath}, C., {Banerjee}, R., \& {K{\"o}rtgen}, B. 2024, \aap, 687, A276

\bibitem[{{Binney} \& {Tremaine}(1987)}]{binney_tremaine_1987_galactic_dynamics}
{Binney}, J. \& {Tremaine}, S. 1987, {Galactic dynamics}

\bibitem[{{Chepurnov} {et~al.}(2010){Chepurnov}, {Lazarian}, {Stanimirovi{\'c}}, {Heiles}, \& {Peek}}]{chepurnov_2010}
{Chepurnov}, A., {Lazarian}, A., {Stanimirovi{\'c}}, S., {Heiles}, C., \& {Peek}, J.~E.~G. 2010, \apj, 714, 1398

\bibitem[{{Dobbs} \& {Bonnell}(2006)}]{dobbs_bonnel_2006}
{Dobbs}, C.~L. \& {Bonnell}, I.~A. 2006, \mnras, 367, 873

\bibitem[{{Dubey} {et~al.}(2008){Dubey}, {Fisher}, {Graziani}, {Jordan}, {Lamb}, {Reid}, {Rich}, {Sheeler}, {Townsley}, \& {Weide}}]{dubey_fisher_2008}
{Dubey}, A., {Fisher}, R., {Graziani}, C., {et~al.} 2008, in Astronomical Society of the Pacific Conference Series, Vol. 385, Numerical Modeling of Space Plasma Flows, ed. N.~V. {Pogorelov}, E.~{Audit}, \& G.~P. {Zank}, 145

\bibitem[{{Elmegreen} \& {Elmegreen}(2019)}]{elmegreen_2019}
{Elmegreen}, B.~G. \& {Elmegreen}, D.~M. 2019, \apjs, 245, 14

\bibitem[{{Elmegreen} {et~al.}(2018){Elmegreen}, {Elmegreen}, \& {Efremov}}]{elmegreen_2018}
{Elmegreen}, B.~G., {Elmegreen}, D.~M., \& {Efremov}, Y.~N. 2018, \apj, 863, 59

\bibitem[{{Elmegreen}(1980)}]{elmegreen_1980}
{Elmegreen}, D.~M. 1980, \apj, 242, 528

\bibitem[{{Elmegreen} {et~al.}(2014){Elmegreen}, {Elmegreen}, {Erroz-Ferrer}, {Knapen}, {Teich}, {Popinchalk}, {Athanassoula}, {Bosma}, {Comer{\'o}n}, {Efremov}, {Gadotti}, {Gil de Paz}, {Hinz}, {Ho}, {Holwerda}, {Kim}, {Laine}, {Laurikainen}, {Men{\'e}ndez-Delmestre}, {Mizusawa}, {Mu{\~n}oz-Mateos}, {Regan}, {Salo}, {Seibert}, \& {Sheth}}]{elmegreen_2014}
{Elmegreen}, D.~M., {Elmegreen}, B.~G., {Erroz-Ferrer}, S., {et~al.} 2014, \apj, 780, 32

\bibitem[{{Elmegreen} {et~al.}(2006){Elmegreen}, {Elmegreen}, {Kaufman}, {Sheth}, {Struck}, {Thomasson}, \& {Brinks}}]{elmegreen_2006}
{Elmegreen}, D.~M., {Elmegreen}, B.~G., {Kaufman}, M., {et~al.} 2006, \apj, 642, 158

\bibitem[{{Federrath} {et~al.}(2010){Federrath}, {Roman-Duval}, {Klessen}, {Schmidt}, \& {Mac Low}}]{FederrathDuvalKlessenSchmidtMacLow2010}
{Federrath}, C., {Roman-Duval}, J., {Klessen}, R.~S., {Schmidt}, W., \& {Mac Low}, M. 2010, \aap, 512, A81

\bibitem[{{Federrath} {et~al.}(2022){Federrath}, {Roman-Duval}, {Klessen}, {Schmidt}, \& {Mac Low}}]{FederrathEtAl2022ascl}
{Federrath}, C., {Roman-Duval}, J., {Klessen}, R.~S., {Schmidt}, W., \& {Mac Low}, M.~M. 2022, {TG: Turbulence Generator}, Astrophysics Source Code Library, record ascl:2204.001

\bibitem[{{Federrath} {et~al.}(2011){Federrath}, {Sur}, {Schleicher}, {Banerjee}, \& {Klessen}}]{FederrathSurSchleicherBanerjeeKlessen2011}
{Federrath}, C., {Sur}, S., {Schleicher}, D.~R.~G., {Banerjee}, R., \& {Klessen}, R.~S. 2011, \apj, 731, 62

\bibitem[{{Fryxell} {et~al.}(2000){Fryxell}, {Olson}, {Ricker}, {Timmes}, {Zingale}, {Lamb}, {MacNeice}, {Rosner}, {Truran}, \& {Tufo}}]{FryxellEtAl2000}
{Fryxell}, B., {Olson}, K., {Ricker}, P., {et~al.} 2000, \apjs, 131, 273

\bibitem[{{Fuchs}(2001)}]{fuchs_2001}
{Fuchs}, B. 2001, \aap, 368, 107

\bibitem[{{Griv} \& {Gedalin}(2012)}]{griv_stability_2012}
{Griv}, E. \& {Gedalin}, M. 2012, \mnras, 422, 600

\bibitem[{{Griv} \& {Wang}(2014)}]{griv_wang_2014_hydro_sims_twoD}
{Griv}, E. \& {Wang}, H.-H. 2014, \na, 30, 8

\bibitem[{{Jeans}(1902)}]{Jeans1902}
{Jeans}, J.~H. 1902, Royal Society of London Philosophical Transactions Series A, 199, 1

\bibitem[{{Jeffreson} {et~al.}(2020){Jeffreson}, {Kruijssen}, {Keller}, {Chevance}, \& {Glover}}]{jeffreson_2020}
{Jeffreson}, S. M.~R., {Kruijssen}, J.~M.~D., {Keller}, B.~W., {Chevance}, M., \& {Glover}, S. C.~O. 2020, \mnras, 498, 385

\bibitem[{{Khullar} {et~al.}(2024){Khullar}, {Matzner}, {Murray}, {Grudi{\'c}}, {Guszejnov}, {Wetzel}, \& {Hopkins}}]{khullar_2024}
{Khullar}, S., {Matzner}, C.~D., {Murray}, N., {et~al.} 2024, \apj, 973, 40

\bibitem[{{Kim} {et~al.}(2016){Kim}, {Agertz}, {Teyssier}, {Butler}, {Ceverino}, {Choi}, {Feldmann}, {Keller}, {Lupi}, {Quinn}, {Revaz}, {Wallace}, {Gnedin}, {Leitner}, {Shen}, {Smith}, {Thompson}, {Turk}, {Abel}, {Arraki}, {Benincasa}, {Chakrabarti}, {DeGraf}, {Dekel}, {Goldbaum}, {Hopkins}, {Hummels}, {Klypin}, {Li}, {Madau}, {Mandelker}, {Mayer}, {Nagamine}, {Nickerson}, {O'Shea}, {Primack}, {Roca-F{\`a}brega}, {Semenov}, {Shimizu}, {Simpson}, {Todoroki}, {Wadsley}, {Wise}, \& {AGORA Collaboration}}]{agora_project_kim_2016}
{Kim}, J.-h., {Agertz}, O., {Teyssier}, R., {et~al.} 2016, \apj, 833, 202

\bibitem[{{Kim} {et~al.}(2020){Kim}, {Kim}, \& {Ostriker}}]{kim_wong_tigress_2021}
{Kim}, W.-T., {Kim}, C.-G., \& {Ostriker}, E.~C. 2020, \apj, 898, 35

\bibitem[{{Kim} {et~al.}(2014){Kim}, {Kim}, \& {Kim}}]{kim_wong_kim_2014}
{Kim}, W.-T., {Kim}, Y., \& {Kim}, J.-G. 2014, \apj, 789, 68

\bibitem[{{Kim} \& {Ostriker}(2002)}]{kim_ostriker_2002}
{Kim}, W.-T. \& {Ostriker}, E.~C. 2002, \apj, 570, 132

\bibitem[{{Kim} \& {Ostriker}(2006)}]{kim_formation_2006}
{Kim}, W.-T. \& {Ostriker}, E.~C. 2006, \apj, 646, 213

\bibitem[{{Kim} {et~al.}(2015){Kim}, {Kim}, \& {Elmegreen}}]{kim_wong_kim_2015}
{Kim}, Y., {Kim}, W.-T., \& {Elmegreen}, B.~G. 2015, \apj, 809, 33

\bibitem[{{K{\"o}rtgen} {et~al.}(2019){K{\"o}rtgen}, {Banerjee}, {Pudritz}, \& {Schmidt}}]{bastian_2019}
{K{\"o}rtgen}, B., {Banerjee}, R., {Pudritz}, R.~E., \& {Schmidt}, W. 2019, \mnras, 489, 5004

\bibitem[{{Kuhn} {et~al.}(2021){Kuhn}, {Benjamin}, {Zucker}, {Krone-Martins}, {de Souza}, {Castro-Ginard}, {Ishida}, {Povich}, \& {Hillenbrand}}]{kuhn_2021}
{Kuhn}, M.~A., {Benjamin}, R.~A., {Zucker}, C., {et~al.} 2021, \aap, 651, L10

\bibitem[{{La Vigne} {et~al.}(2006){La Vigne}, {Vogel}, \& {Ostriker}}]{vigne_2006}
{La Vigne}, M.~A., {Vogel}, S.~N., \& {Ostriker}, E.~C. 2006, \apj, 650, 818

\bibitem[{{Lang} {et~al.}(2020){Lang}, {Meidt}, {Rosolowsky}, {Nofech}, {Schinnerer}, {Leroy}, {Emsellem}, {Pessa}, {Glover}, {Groves}, {Hughes}, {Kruijssen}, {Querejeta}, {Schruba}, {Bigiel}, {Blanc}, {Chevance}, {Colombo}, {Faesi}, {Henshaw}, {Herrera}, {Liu}, {Pety}, {Puschnig}, {Saito}, {Sun}, \& {Usero}}]{lang_phangs_rotation_2020}
{Lang}, P., {Meidt}, S.~E., {Rosolowsky}, E., {et~al.} 2020, \apj, 897, 122

\bibitem[{{Lee}(2014)}]{lee_feathering_2014}
{Lee}, W.-K. 2014, \apj, 792, 122

\bibitem[{{Lee} \& {Shu}(2012)}]{lee_feathering_2012}
{Lee}, W.-K. \& {Shu}, F.~H. 2012, \apj, 756, 45

\bibitem[{{Leroy} {et~al.}(2019){Leroy}, {Sandstrom}, {Lang}, {Lewis}, {Salim}, {Behrens}, {Chastenet}, {Chiang}, {Gallagher}, {Kessler}, \& {Utomo}}]{leroy_2019_galex_wise}
{Leroy}, A.~K., {Sandstrom}, K.~M., {Lang}, D., {et~al.} 2019, \apjs, 244, 24

\bibitem[{{Leroy} {et~al.}(2021){Leroy}, {Schinnerer}, {Hughes}, {Rosolowsky}, {Pety}, {Schruba}, {Usero}, {Blanc}, {Chevance}, {Emsellem}, {Faesi}, {Herrera}, {Liu}, {Meidt}, {Querejeta}, {Saito}, {Sandstrom}, {Sun}, {Williams}, {Anand}, {Barnes}, {Behrens}, {Belfiore}, {Benincasa}, {Be{\v{s}}li{\'c}}, {Bigiel}, {Bolatto}, {den Brok}, {Cao}, {Chandar}, {Chastenet}, {Chiang}, {Congiu}, {Dale}, {Deger}, {Eibensteiner}, {Egorov}, {Garc{\'\i}a-Rodr{\'\i}guez}, {Glover}, {Grasha}, {Henshaw}, {Ho}, {Kepley}, {Kim}, {Klessen}, {Kreckel}, {Koch}, {Kruijssen}, {Larson}, {Lee}, {Lopez}, {Machado}, {Mayker}, {McElroy}, {Murphy}, {Ostriker}, {Pan}, {Pessa}, {Puschnig}, {Razza}, {S{\'a}nchez-Bl{\'a}zquez}, {Santoro}, {Sardone}, {Scheuermann}, {Sliwa}, {Sormani}, {Stuber}, {Thilker}, {Turner}, {Utomo}, {Watkins}, \& {Whitmore}}]{leroy_phangs_co_2021}
{Leroy}, A.~K., {Schinnerer}, E., {Hughes}, A., {et~al.} 2021, \apjs, 257, 43

\bibitem[{{Levy} {et~al.}(2018){Levy}, {Bolatto}, {Teuben}, {S{\'a}nchez}, {Barrera-Ballesteros}, {Blitz}, {Colombo}, {Garc{\'\i}a-Benito}, {Herrera-Camus}, {Husemann}, {Kalinova}, {Lan}, {Leung}, {Mast}, {Utomo}, {van de Ven}, {Vogel}, \& {Wong}}]{levy_edge-califa_2018}
{Levy}, R.~C., {Bolatto}, A.~D., {Teuben}, P., {et~al.} 2018, \apj, 860, 92

\bibitem[{{Mandowara} {et~al.}(2022){Mandowara}, {Sormani}, {Sobacchi}, \& {Klessen}}]{mandowara_physical_2022}
{Mandowara}, Y., {Sormani}, M.~C., {Sobacchi}, E., \& {Klessen}, R.~S. 2022, \mnras, 513, 5052

\bibitem[{{McClure-Griffiths} {et~al.}(2023){McClure-Griffiths}, {Stanimirovi{\'c}}, \& {Rybarczyk}}]{naomi_2023_hi}
{McClure-Griffiths}, N.~M., {Stanimirovi{\'c}}, S., \& {Rybarczyk}, D.~R. 2023, \araa, 61, 19

\bibitem[{{McGaugh} \& {Schombert}(2015)}]{mcGaugh_schombert_2015}
{McGaugh}, S.~S. \& {Schombert}, J.~M. 2015, \apj, 802, 18

\bibitem[{{Meidt}(2022)}]{meidt_molecular_2022}
{Meidt}, S.~E. 2022, \apj, 937, 88

\bibitem[{{Meidt} {et~al.}(2018){Meidt}, {Leroy}, {Rosolowsky}, {Kruijssen}, {Schinnerer}, {Schruba}, {Pety}, {Blanc}, {Bigiel}, {Chevance}, {Hughes}, {Querejeta}, \& {Usero}}]{meidt_2018}
{Meidt}, S.~E., {Leroy}, A.~K., {Rosolowsky}, E., {et~al.} 2018, \apj, 854, 100

\bibitem[{{Meidt} {et~al.}(2023){Meidt}, {Rosolowsky}, {Sun}, {Koch}, {Klessen}, {Leroy}, {Schinnerer}, {Barnes}, {Glover}, {Lee}, {van der Wel}, {Watkins}, {Williams}, {Bigiel}, {Boquien}, {Blanc}, {Cao}, {Chevance}, {Dale}, {Egorov}, {Emsellem}, {Grasha}, {Henshaw}, {Kruijssen}, {Larson}, {Liu}, {Murphy}, {Pety}, {Querejeta}, {Saito}, {Sandstrom}, {Smith}, {Sormani}, \& {Thilker}}]{meidt_rosolowsky_23}
{Meidt}, S.~E., {Rosolowsky}, E., {Sun}, J., {et~al.} 2023, \apjl, 944, L18

\bibitem[{{Meidt} \& {van der Wel}(2024)}]{meidt_2024}
{Meidt}, S.~E. \& {van der Wel}, A. 2024, \apj, 966, 62

\bibitem[{{Pantaleoni Gonz{\'a}lez} {et~al.}(2021){Pantaleoni Gonz{\'a}lez}, {Ma{\'\i}z Apell{\'a}niz}, {Barb{\'a}}, \& {Reed}}]{pantaleoni_2021}
{Pantaleoni Gonz{\'a}lez}, M., {Ma{\'\i}z Apell{\'a}niz}, J., {Barb{\'a}}, R.~H., \& {Reed}, B.~C. 2021, \mnras, 504, 2968

\bibitem[{{Parker}(1966)}]{parker_1966}
{Parker}, E.~N. 1966, \apj, 145, 811

\bibitem[{{Puerari} {et~al.}(2014){Puerari}, {Elmegreen}, \& {Block}}]{puerari_2014}
{Puerari}, I., {Elmegreen}, B.~G., \& {Block}, D.~L. 2014, \aj, 148, 133

\bibitem[{{Robinson} \& {Wadsley}(2024)}]{robinson_wadsley_2024}
{Robinson}, H. \& {Wadsley}, J. 2024, \mnras, 534, 1420

\bibitem[{{Romeo} \& {Falstad}(2013)}]{romeo_falstad_2013}
{Romeo}, A.~B. \& {Falstad}, N. 2013, \mnras, 433, 1389

\bibitem[{{Romeo} \& {Wiegert}(2011)}]{romeo_wiegert_2011_q_stability}
{Romeo}, A.~B. \& {Wiegert}, J. 2011, \mnras, 416, 1191

\bibitem[{{Savchenko}(2012)}]{savchenko_2012}
{Savchenko}, S.~S. 2012, Astrophysical Bulletin, 67, 310

\bibitem[{{Schinnerer} {et~al.}(2017){Schinnerer}, {Meidt}, {Colombo}, {Chandar}, {Dobbs}, {Garc{\'\i}a-Burillo}, {Hughes}, {Leroy}, {Pety}, {Querejeta}, {Kramer}, \& {Schuster}}]{schinnerer_pdbi_2017}
{Schinnerer}, E., {Meidt}, S.~E., {Colombo}, D., {et~al.} 2017, \apj, 836, 62

\bibitem[{{Shetty} \& {Ostriker}(2006)}]{shetty_ostriker_2006}
{Shetty}, R. \& {Ostriker}, E.~C. 2006, \apj, 647, 997

\bibitem[{{Sormani} {et~al.}(2017){Sormani}, {Sobacchi}, {Shore}, {Tre{\ss}}, \& {Klessen}}]{sormani_sobacchi_2017}
{Sormani}, M.~C., {Sobacchi}, E., {Shore}, S.~N., {Tre{\ss}}, R.~G., \& {Klessen}, R.~S. 2017, \mnras, 471, 2932

\bibitem[{{Thilker} {et~al.}(2023){Thilker}, {Lee}, {Deger}, {Barnes}, {Bigiel}, {Boquien}, {Cao}, {Chevance}, {Dale}, {Egorov}, {Glover}, {Grasha}, {Henshaw}, {Klessen}, {Koch}, {Kruijssen}, {Leroy}, {Lessing}, {Meidt}, {Pinna}, {Querejeta}, {Rosolowsky}, {Sandstrom}, {Schinnerer}, {Smith}, {Watkins}, {Williams}, {Anand}, {Belfiore}, {Blanc}, {Chandar}, {Congiu}, {Emsellem}, {Groves}, {Kreckel}, {Larson}, {Liu}, {Pessa}, \& {Whitmore}}]{thilker_filamentary_NGC628_JWST_2023}
{Thilker}, D.~A., {Lee}, J.~C., {Deger}, S., {et~al.} 2023, \apjl, 944, L13

\bibitem[{{Toomre}(1964)}]{toomre_1964}
{Toomre}, A. 1964, \apj, 139, 1217

\bibitem[{{Tress} {et~al.}(2020){Tress}, {Sormani}, {Glover}, {Klessen}, {Battersby}, {Clark}, {Hatchfield}, \& {Smith}}]{tress_simulations_2020}
{Tress}, R.~G., {Sormani}, M.~C., {Glover}, S. C.~O., {et~al.} 2020, \mnras, 499, 4455

\bibitem[{{Truelove} {et~al.}(1997){Truelove}, {Klein}, {McKee}, {Holliman}, {Howell}, \& {Greenough}}]{TrueloveEtAl1997}
{Truelove}, J.~K., {Klein}, R.~I., {McKee}, C.~F., {et~al.} 1997, \apjl, 489, L179

\bibitem[{{Veena} {et~al.}(2021){Veena}, {Schilke}, {S{\'a}nchez-Monge}, {Sormani}, {Klessen}, {Schuller}, {Colombo}, {Csengeri}, {Mattern}, \& {Urquhart}}]{veena_2021}
{Veena}, V.~S., {Schilke}, P., {S{\'a}nchez-Monge}, {\'A}., {et~al.} 2021, \apjl, 921, L42

\bibitem[{{Wada} \& {Koda}(2004)}]{wada_koda_2004}
{Wada}, K. \& {Koda}, J. 2004, \mnras, 349, 270

\bibitem[{{Wang} {et~al.}(2010){Wang}, {Klessen}, {Dullemond}, {van den Bosch}, \& {Fuchs}}]{wang_equilibrium_2010}
{Wang}, H.-H., {Klessen}, R.~S., {Dullemond}, C.~P., {van den Bosch}, F.~C., \& {Fuchs}, B. 2010, \mnras, 407, 705

\bibitem[{{Williams} {et~al.}(2022){Williams}, {Sun}, {Barnes}, {Schinnerer}, {Henshaw}, {Meidt}, {Querejeta}, {Watkins}, {Bigiel}, {Blanc}, {Boquien}, {Cao}, {Chevance}, {Egorov}, {Emsellem}, {Glover}, {Grasha}, {Hassani}, {Jeffreson}, {Jim{\'e}nez-Donaire}, {Kim}, {Klessen}, {Kreckel}, {Kruijssen}, {Larson}, {Leroy}, {Liu}, {Pessa}, {Pety}, {Pinna}, {Rosolowsky}, {Sandstrom}, {Smith}, {Sormani}, {Stuber}, {Thilker}, \& {Whitmore}}]{williams_phangs-jwst_2022}
{Williams}, T.~G., {Sun}, J., {Barnes}, A.~T., {et~al.} 2022, \apjl, 941, L27

\bibitem[{{Zhao} {et~al.}(2024){Zhao}, {Pudritz}, {Pillsworth}, {Robinson}, \& {Wadsley}}]{bo_pudritz_2024}
{Zhao}, B., {Pudritz}, R.~E., {Pillsworth}, R., {Robinson}, H., \& {Wadsley}, J. 2024, \apj, arXiv:2405.18474

\bibitem[{{Zucker} {et~al.}(2015){Zucker}, {Battersby}, \& {Goodman}}]{zucker_bones_of_MW_2015}
{Zucker}, C., {Battersby}, C., \& {Goodman}, A. 2015, \apj, 815, 23

\end{thebibliography}

\appendix

\section{Dark matter potential} \label{appendix:dark_matter_potential}

Here, we determine the stellar plus dark matter potential $\phi_\mathrm{dm}$ such that our disc is in radial equilibrium. In a reference frame that co-rotates with the disc angular frequency $\Omega$ at a given position $(R,z)$, radial equilibrium demands that the sum of the gravitational, pressure, and centrifugal forces vanish. We can express this condition as 
\begin{equation}
    \rho \frac{\partial}{\partial R}\left(\phi_\mathrm{dm} + \phi_\mathrm{g}\right) + \cs^2 \frac{\partial \rho}{\partial R} - \rho \Omega^2 R = 0,
\end{equation}
where $\phi_\mathrm{g}$ is the potential due to disc self-gravity, which is given by \citet[][their equation 29]{wang_equilibrium_2010}
\begin{equation}
    \phi_\mathrm{g} = 4\pi G \Sigma_\circ \rd y^2 \left[I_0(y) K_0(y) - I_1(y) K_1(y)\right],
\end{equation}
where $y = R/2\rd$ and $I_0$, $K_0$, $I_1$, and $K_1$ are the modified Bessel functions of the first and second kind, of order zero and one.\footnote{Formally the expression that \citet{wang_equilibrium_2010} provide is for a purely exponential disc, without the central flattening we have introduced. However, since our disc differs in mass from a pure exponential disc by only $\approx 5\%$, we expect the difference in potential to be small.} To proceed we must now make some assumptions about the vertical structure of the disc and potential. We start by assuming that the radial variation in density is approximately exponential, $\rho(R) \propto \exp(-R/R_\mathrm{d})$. This assumption is exact only if the disc vertical profile is independent of radius, but is true approximately as long as the disc scale height does not vary on scales $\ll R_\mathrm{d}$; since this term is highly sub-dominant in any event (roughly $100\times$ smaller than the centrifugal term), we do not seek a more accurate approximation. Finally, at the midplane we require that $\Omega = v_\mathrm{rot}/R$, but we must make an assumption about the variation of $\Omega$ in the $z$ direction, or, equivalently, assume a shape for the centrifugal potential $\psi$, where $\partial\psi/\partial R = \Omega^2 R$. For this purpose we adopt the flattened centrifugal potential form suggested by \citet{binney_tremaine_1987_galactic_dynamics},
\begin{equation} \label{eqn:psi_dm_and_stellar}
    \psi = \frac{\vc ^{2}}{2} \ln \left \{ \frac{1}{R_{\rm c} ^{2}} \left [ R^2 + R_{\rm c}^2 + \left ( \frac{z}{q} \right )^2 \right] \right \},
\end{equation}
where $q$ is a dimensionless parameter that determines the amount of flattening. One can readily verify that this potential satisfies our condition that $\partial\psi/\partial R = \Omega^2 R$. Following \citeauthor{binney_tremaine_1987_galactic_dynamics}'s estimates for realistic galactic potentials, we adopt $q = 0.7$. With these assumptions, our equation of force balance reduces to
\begin{equation}
    \rho \frac{\partial}{\partial R}\left(\phi_\mathrm{dm} + \phi_\mathrm{g} - \cs^2 \frac{R}{R_d} - \psi \right) = 0,
\end{equation}
which has the immediate solution
\begin{equation} \label{eqn:phi_dm_analytical_form}
    \phi_{\rm dm} = \psi - \phi_{\rm g} + \cs^2 \frac{R}{\rd}.
\end{equation}
We have therefore succeeded in identifying the stellar plus dark matter potential required to place the disc in equilibrium.

\section{Two dimensionless parameters} \label{appendix:two_dimensionless_parameters}

In this appendix we prove the assertion made in the main text that the fluid equations that we solve depend only upon the two dimensionless parameters $\langle Q\rangle$ and $\machc$. And that these together with the additional dimensionless parameters describing the initial conditions that we do not vary -- $\rc/\rd$, $R_\mathrm{max}/\rd$, $q$, $\alpha$, and $\beta$ -- fully determine the evolution of the system. To do so, we non-dimensionalise the fluid variables via the transformations 
\begin{align}
    \vec{r} &= \vec{x} \rd, \\
    t &= \tau \frac{\rd}{\cs}, \\
    \vec{v} &= \vec{u}\cs,  \\
    \rho &= a\frac{\Sigma_{\circ}}{\rd}, \\
    \Tilde{\phi} &= \frac{\phi}{\cs^{2}}.
\end{align}
where $x$, $\tau$, $u$, $a$, and $\Tilde{\phi}$ are dimensionless length, time, velocity, density and the gravitational potential, respectively, scaled by the appropriate combinations of the disc scale length $\rd$, the sound speed $\cs$ and the surface density scaling of our galaxies $\Sigma_{\circ}$. The equations of mass and momentum conservation plus the Poisson equation under this transformation are
\begin{align}
    \frac{\partial a}{\partial \tau} + \frac{1}{\machc} \vec{\nabla_{x}}(a\vec{u})    =&~0, \label{eqn:dimensionless_cons_of_mass} \\
    \machc\frac{\partial \vec{u} }{\partial \tau} + (\vec{u} \cdot \nabla_{x}) \cdot \vec{u} =& -\frac{\nabla_{x} a}{a} - \nabla_{x} \left  (\Tilde{\phi}_{\rm g} + \Tilde{\phi}_{\rm dm} \right ), \label{eqn:dimensionless_cons_of_momentum} \\
 \nabla^{2}_{x} \Tilde{\phi} =& \xi \frac{\machc}{\langle Q\rangle} a \label{eqn:dimensionless_poison},
\end{align}
where $\nabla_x$ indicates gradient with respect to the dimensionless position variable $\mathbf{x}$ and $\xi$ is a dimensionless constant given by
\begin{eqnarray}
    \xi &= &
    \frac{4 \sqrt{2}}{X_\mathrm{max}} \int_0^{X_\mathrm{max}} \frac{1}{\exp\left[-X-\beta\exp\left(-\alpha X\right)\right]} 
   \frac{\sqrt{X^2+2X_\mathrm{c}^2}}{X^2 + X_\mathrm{c}^2}
    \, dX 
    \nonumber \\
    & \approx & 19.43.
\end{eqnarray}
In the expression above, we have defined $X = R/\rd$ as the dimensionless cylindrical radius, $X_\mathrm{c}$ and $X_\mathrm{max}$ are defined analogously, and the numerical evaluation is for our fixed values of these quantities, $\alpha$, and $\beta$. Note that we have used the isothermal equation of state to substitute for the pressure (\autoref{eqn:isothermal_eos}). Finally, we can rewrite the various components of the gravitational potential that contributed to $\Tilde{\phi}$ using the same non-dimensionalisation procedure. Substituting into \autoref{eqn:phi_dm_analytical_form} we have
\begin{align}
    \Tilde{\phi}_{\rm dm} &= \Tilde{\psi} - \Tilde{\phi}_{\rm g} + X, \\
    \Tilde{\psi} &= \frac{\machc^{2}}{2} \ln \left \{ X_\mathrm{c}^{-2} \left [ X^2 + X_\mathrm{c}^2 + \left ( \frac{Z}{q} \right )^2 \right] \right \}, \\
    \Tilde{\phi}_{\rm g} &= \frac{\machc}{\langle Q\rangle \xi} X^2 \left [ I_{0} \left(\frac{X}{2}\right) K_{0} \left(\frac{X}{2}\right) -  I_{1} \left(\frac{X}{2}\right) K_{1}\left(\frac{X}{2}\right)\right ],
\end{align}
where $Z = z/\rd$ is the dimensionless vertical height. This completes the non-dimensionalisation of the system, and demonstrates that the results depend only on $\langle Q\rangle$, $\machc$, $X_\mathrm{c}$, $X_\mathrm{max}$, $q$, $\alpha$, and $\beta$ as claimed.

\section{Toomre $\langle Q \rangle =2$ run}\label{appendix:Q2_run}

\begin{figure*}
    \includegraphics[width=\linewidth]{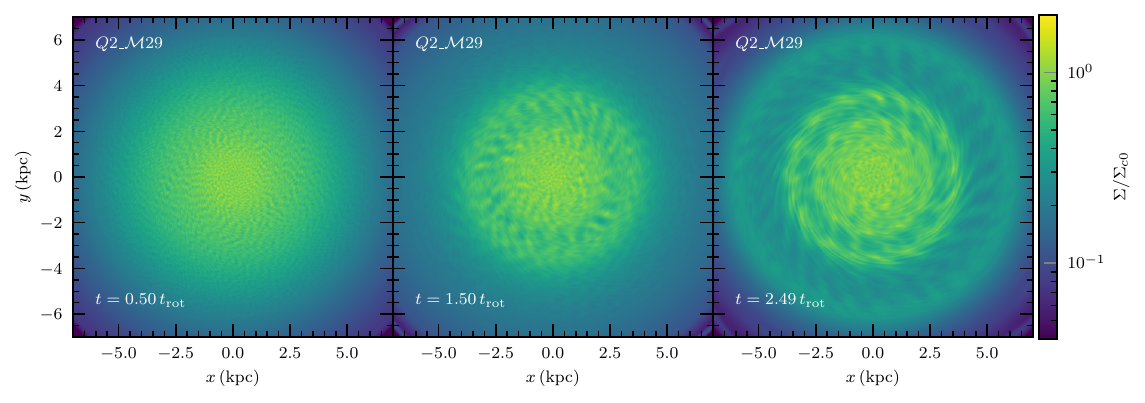}
    \caption{Same as \autoref{fig:projection_Q1_mach29_evolution}, but for the longer time evolution of the $Q=2$ run. We see that there are faint structures at early times in the first two panels. The third panel shows a structure with somewhat more density contrast, visually confirming the slow growth of an instability.}
    \label{fig:projectionPanel_Q2}
\end{figure*}
In this Appendix we present the long-term evolution of the $\langle Q \rangle =2$ run ($Q2\_\mach29$), which does not produce significant structure on the same $\sim t_\mathrm{rot}$ timescales as the $\langle Q\rangle = 1$ runs presented in the main body. We continue this run to time $t= 2.5\,t_\mathrm{rot}$, and show the projected surface density as a function of time in \autoref{fig:projectionPanel_Q2}. We see that the galaxy does develop structures with noticeable density contrast by $t = 2.5 \, \trot$ at $R=3\, \si{kpc}$, and that as in the $\langle Q\rangle = 1$ cases, these structures are regularly spaced in the azimuth. However, there are also significant differences. First, the structures in the $\langle Q\rangle =2$ run clearly have lower pitch angles and density contrasts. Second, their azimuthal spacing is also different - counting the number of dense structures along the azimuth we estimate $m_{\rm fil}\simeq15$ to be the dominant mode of the filaments. This is a factor of two less than the $Q1\_\mach29$ run, that has the same $\machc$ value as the $\langle Q \rangle = 2$ run. Moreover, the number of filaments in the $Q2\_\mach29$ case is around the value expected from the swing amplification mechanism, which is $=13$.

\begin{figure}
    \includegraphics[width=0.97\linewidth]{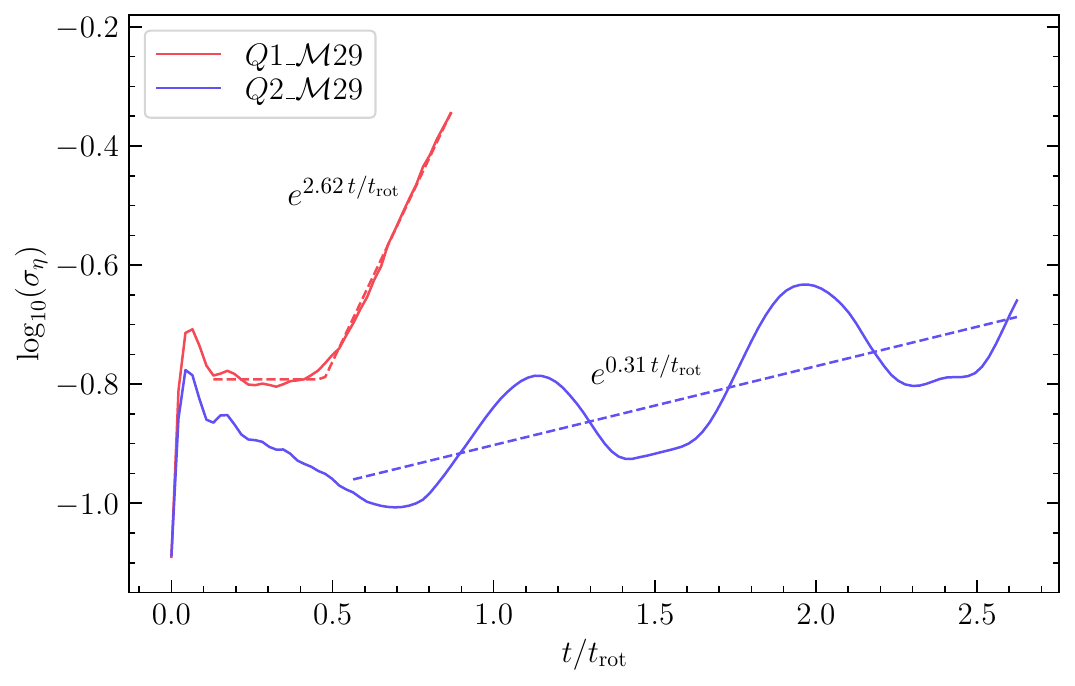}
    \caption{Same as \autoref{fig:clumpingFactor_timeEvol}, but only for $Q1\_\mach29$ and $Q2\_\mach29$ runs, where the latter was evolved to a later time than shown in the main text. We see the oscillatory peaks in the $Q=2$ run, but also a steady, yet slow growth over time. This growth is visibly smaller than in the $Q1\_\mach29$ run.}
    \label{fig:clumpingFactor_timeEvol_Q2_Q1}
\end{figure}

To compare the growth rate of filaments of both the cases, we plot the time evolution of the variation of the logarithmic density contrast ($\sigma_{\eta}$) of the $Q1\_\mach29$ run and the $Q2\_\mach29$ run together in \autoref{fig:clumpingFactor_timeEvol_Q2_Q1}. This is similar to the left panel of the \autoref{fig:clumpingFactor_timeEvol}, except for the x-axis that extends to longer times. The dashed lines show the empirical fit on the two runs. The $Q1\_\mach29$ run has the piecewise-fit of the form \autoref{eqn: piece-wise_fit}, identical to the one shown in the right panel of the \autoref{fig:clumpingFactor_timeEvol}. We fit only the exponential part of the fit for the $Q2\_\mach29$ run after $t = 0.55\, \trot$, since we are interested exclusively in the rising part of the curve. As seen in the main text, the $Q2\_\mach29$ shows an initial decline in $\sigma_{\eta}$. However, from $t \approx 0.55$ onwards, it shows periodic oscillations with a rising mean. From our empirical fit to this part of the curve, we find $\omega t_\mathrm{rot} = 0.31$, which is $\approx 8$ times smaller than $Q1\_\mach29$ case.

\section{Resolution study} \label{appendix:resolutionStudy}

\begin{figure}[!h]
    \includegraphics[width=0.97\linewidth]{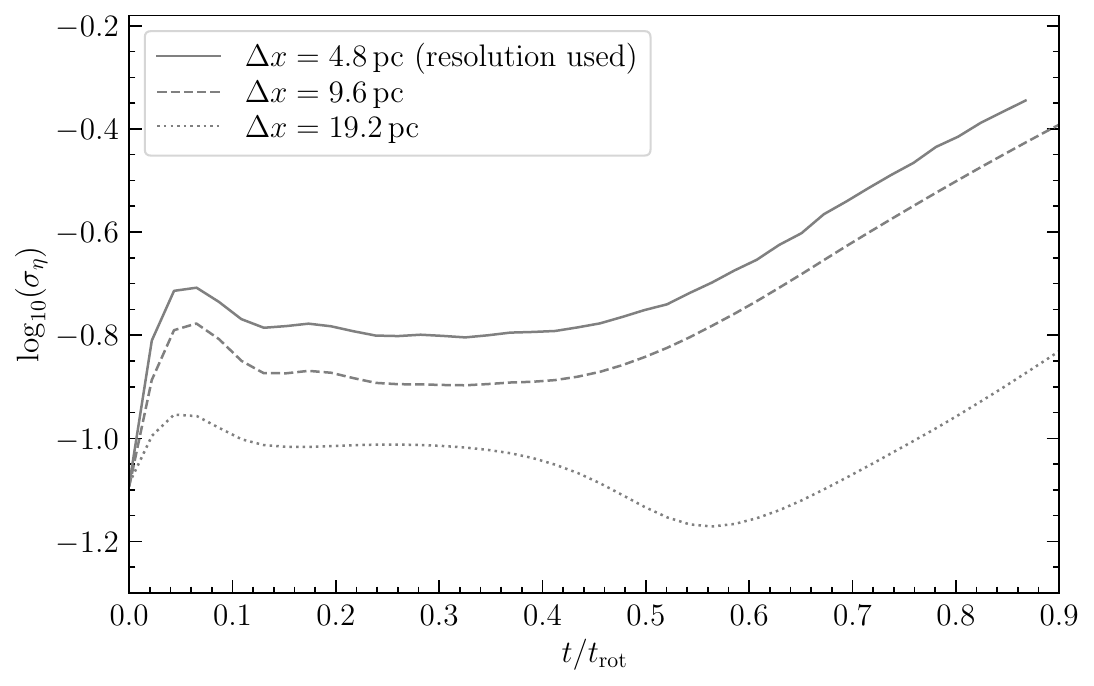}
    \caption{Same as \autoref{fig:clumpingFactor_timeEvol}, but for the $Q1\_\mach29$ run at three different resolutions. The solid line depicts the resolution used in the main text ($\Delta x = 4.8\, \si{pc}$) and the other two lines are runs with resolutions reduced by a factor of two each. }
    \label{fig:resolution_study}
\end{figure}

For a resolution study, we perform two runs with resolutions that are a factor of two ($\Delta x = 9.6\, \si{pc}$) and four ($\Delta x = 19.2\, \si{pc}$) lower than the one used in the main text. Similar to \autoref{fig:clumpingFactor_timeEvol}, we show the time evolution of the standard deviation of the logarithmic surface density (defined in \autoref{eqn:clumping_factor}) of the annuli $R = 2-4\, \si{kpc}$ of the three runs in \autoref{fig:resolution_study}. We see similar qualitative evolution in all the three cases as seen for all the $\langle Q \rangle =1$ runs. The steady value of $\sigma_{\eta}$ after the initial relaxation relaxation phase increases with increasing resolution. This is expected since we resolve the initial turbulent velocity fluctuations and the resulting density fluctuations with more cells. However, this change in $\sigma_{\eta}$ between runs with consecutive increasing resolutions is less than a factor of two. The time of the onset of exponential growth, as well as the slopes of the lines following the onset are very similar, especially for the two runs with the highest resolutions.

\end{document}